\documentclass[letterpaper,twocolumn,10pt]{article}
\usepackage{usenix-2020-09}
\usepackage[available,functional]{usenixbadges}

\usepackage{tikz}
\usepackage{amsmath}
\usepackage{authblk}

\usepackage{filecontents}
\usepackage{booktabs}
\usepackage[dvipsnames]{xcolor}

\usepackage{listings}
\definecolor{delim}{RGB}{20,105,176}
\definecolor{numb}{RGB}{106, 109, 32}
\definecolor{string}{rgb}{0.64,0.08,0.08}
\lstdefinelanguage{json}{
    showspaces=false,
    showtabs=false,
    breaklines=true,
    postbreak=\raisebox{0ex}[0ex][0ex]{\ensuremath{\color{gray}\hookrightarrow\space}},
    breakatwhitespace=true,
    basicstyle=\ttfamily\small,
    upquote=true,
    morestring=[b]",
    stringstyle=\color{string},
    literate=
     *{0}{{{\color{numb}0}}}{1}
      {1}{{{\color{numb}1}}}{1}
      {2}{{{\color{numb}2}}}{1}
      {3}{{{\color{numb}3}}}{1}
      {4}{{{\color{numb}4}}}{1}
      {5}{{{\color{numb}5}}}{1}
      {6}{{{\color{numb}6}}}{1}
      {7}{{{\color{numb}7}}}{1}
      {8}{{{\color{numb}8}}}{1}
      {9}{{{\color{numb}9}}}{1}
      {\{}{{{\color{delim}{\{}}}}{1}
      {\}}{{{\color{delim}{\}}}}}{1}
      {[}{{{\color{delim}{[}}}}{1}
      {]}{{{\color{delim}{]}}}}{1},
}

\usepackage{tcolorbox}
\usepackage{enumitem}
\newcommand{\roundframe}[1]{{\setlength\fboxrule{0pt}\fbox{\tcbox[colframe=black,colback=white,shrink tight,boxrule=0.5pt,extrude by=2.5pt]{\small #1}}}}

\newcommand{\parvspace}{\vspace{0.20cm}}

\RequirePackage{fontawesome}
\DeclareRobustCommand{\clockMessageSending}{\text{\reflectbox{\faClockO}}}

\usepackage[flushmargin]{footmisc}

\usepackage{tcolorbox,url}
\tcbuselibrary{skins,xparse,listings}
\lstdefinestyle{Bash}{
  basicstyle=\small\ttfamily,
  numbers=none,
  frame=tblr,
  columns=fullflexible,
  backgroundcolor=\color{blue!10},
  linewidth=0.99\linewidth,
  xleftmargin=0.01\linewidth
}

\newtcblisting{myTerminal}{colback=violet!50!black,
colupper=white,colframe=gray!65!black,listing only,
top=1.5pt,bottom=1.5pt,left=1.5pt,right=1pt,
listing options={style=tcblatex,escapeinside=``,},
every listing line={\MyTerminalPrompt}}
\pgfkeys{/myTerminal/.cd,
user/.code={\gdef\MyTerminalUser{#1}},user={},
host/.code={\gdef\MyTerminalHost{#1}},host={},
color/.code={\gdef\MyTerminalColor{#1}},color=white,
prompt char/.code={\gdef\MyTerminalPromptChar{#1}},prompt char=\$,
root/.style={user=prekey,host=pogo,color=lime,prompt char=\$},
}
\newcommand{\terminalSU}[1]{\pgfkeys{/myTerminal/.cd,#1}%
\gdef\MyTerminalPrompt{\textcolor{\MyTerminalColor}{\small\ttfamily\bfseries \MyTerminalUser@\MyTerminalHost{\textcolor{white}:}\textcolor{cyan!60}{\url{~}}{\textcolor{white}\MyTerminalPromptChar} }}}
\newcommand{\StartTerminalConsole}{\gdef\MyTerminalPrompt{}}
\terminalSU{user=pogo,host=prekey,color=lime}

\usepackage{placeins}

\usepackage{stfloats}

\begin{document}

\date{}

\title{\Large \bf
Prekey Pogo:
Investigating Security and Privacy Issues\\in WhatsApp's Handshake Mechanism
}

\makeatletter
\makeatother

\author[1,2]{Gabriel K. Gegenhuber}
\author[3]{Philipp É. Frenzel}
\author[1]{Maximilian Günther}
\author[1]{Aljosha Judmayer}

\affil[1]{University of Vienna, Faculty of Computer Science}
\affil[2]{UniVie Doctoral School Computer Science}
\affil[3]{SBA~Research}

\maketitle

\begin{abstract}
WhatsApp, the world's largest messaging application, uses a version of the Signal protocol to provide end-to-end encryption (E2EE) with strong security guarantees, including Perfect Forward Secrecy (PFS).
To ensure PFS right from the start of a new conversation --even when the recipient is offline-- a stash of ephemeral (one-time) prekeys must be stored on a server. While the critical role of these one-time prekeys in achieving PFS has been outlined in the Signal specification, we are the first to demonstrate a targeted depletion attack against them on individual WhatsApp user devices.
Our findings not only reveal an attack that can degrade PFS for certain messages, but also expose inherent privacy risks and serious availability implications arising from the refilling and distribution procedure essential for this security mechanism. 

\end{abstract}

\section{Introduction}

WhatsApp is the world's largest messaging application, with more than 3 billion users worldwide~\cite{noauthor_whatsapp_about}. 
Under the hood, WhatsApp uses its own version of the Signal protocol for end-to-end encryption (E2EE) of messages~\cite{noauthor_whatsapp_2023}. 

The Signal protocol suite consists of several different protocols~\cite{marlinspike_x3dh_2016, perrin_double_2016, marlinspike_private_2014, marlinspike_sesame_2017, kret_pqxdh_2023} which together form one of the best end-to-end encrypted communication options available to end users today. 
Parts of the protocol suite have also been formally analyzed and proven secure in their respective security models~\cite{cohn-gordon_post-compromise_2016, alwen_double_2019, cohn-gordon_formal_2020, brendel_post-quantum_2022, cremers_formal_2023}. 
Nonetheless, it remains crucial to continuously analyze protocols in their entirety—including their real-world composition and implementations—to uncover and test new attack strategies, identify real-world limitations, and enhance them accordingly.
For our research, we depleted the \emph{ephemeral prekeys} (also \emph{one-time prekeys}) of our test accounts to analyze attacks on \emph{perfect forward secrecy} (PFS) and to highlight novel privacy and availability implications arising from the current replenishing and distribution mechanisms for such prekey bundles. 

The importance of one-time prekeys for the PFS of initial messages has already been noted in the specification of Signal's X3DH protocol~\cite{marlinspike_x3dh_2016}:

\emph{''This reduction in initial forward secrecy could also happen if one party maliciously drains another party's one-time prekeys, so the server should attempt to prevent this, e.g. with rate limits on fetching prekey bundles.''}

To the best of our knowledge, we are not only the first to test this concrete attack against forward secrecy, but also the first to analyze its feasibility and the general implications of this feature regarding the privacy of users. 
Hereby, we not only show that WhatsApp currently does not employ any rate limiting on fetching prekey bundles of participants, but also highlight that the lack of a detailed specification on how to handle and replenish ephemeral one-time prekeys, allows for device fingerprinting and gives away the online status of the targeted device. Moreover, extensively querying prekey bundles for a targeted account may cause errors, potentially preventing the retrieval of \emph{any} prekey bundle for that account (even without one-time prekeys). As a result, no one would be able to establish new chat sessions with the victim, leading to an availability issue.
While PFS is undoubtedly affected as well, we consider the real world confidentiality impact of this attack to be modest. This is due to the careful design and a clever defense-in-depth strategy of the Signal protocol suite. After being able to circumvent the noise protocol~\cite{perrin2018noise}, an attacker would still need to get his hands on the long- and medium-term keys of a victim to exploit the lack of forward secrecy and decrypt the previously recorded messages. Even without one-time prekeys, the ``self-healing'' properties of the double ratchet~\cite{perrin_double_2016} restore forward secrecy after the first round trip. Nevertheless, the attack on PFS shows that the strong claim from the WhatsApp whitepaper, would at least require a footnote that this currently might not hold for all messages:

\emph{''Due to the ephemeral nature of the cryptographic keys, even
in a situation where the current encryption keys from a user’s device are
physically compromised, they cannot be used to decrypt previously transmitted
messages.''}~\cite{noauthor_whatsapp_2023}

\subsection{Threat Model}

We consider two different attack models: A \emph{PFS attack model} where the attacker is assumed to have far-reaching capabilities, as well as a \emph{privacy and availability attack model} where the sole requirement for the attacker is having a WhatsApp account and the phone number of the target.

\subsubsection{PFS Attack Model}
\label{sec:pfs_threat_model}
The goal of the attacker in this case is as follows:
\begin{itemize}
    \item[\roundframe{G1}] \textbf{PFS Degradation and Exploitation:} Force Bob's communication partners to send initial messages without forward secrecy and exploit this by recording the respective messages for possible decryption later on after Bob's long- and medium- term secret keys have been compromised. 
\end{itemize}

\parvspace
\noindent
\textbf{Attacker Capabilities.} In this attack, we assume a passive attacker that has access to WhatsApp's data center internal network communication, s.t., they are able to gain access to end-to-end encrypted messages of WhatsApp users. In other words, the attacker is able to strip the first layer of transport encryption -- usually provided by TLS or the Noise protocol framework between the client and the server~\cite{perrin2018noise} -- on top of the E2EE communication used in the Signal protocol suite. 
This is not an unrealistic scenario if a nation state actor, or compromised WhatsApp inter-server communication (e.g., by an internal employee), is considered. 
Moreover, minimizing the trust in the operator of the servers (i.e., WhatsApp in this case), is an explicit design goal of the Signal protocol family~\cite{private_groups}. It is therefore reasonable to assume that an attacker could, under certain circumstances, gain access to end-to-end encrypted (E2EE) messages.

Moreover, we require the attacker to know the phone number of the target user, as well as a WhatsApp account.

\parvspace
\noindent
\textbf{Prekey Depletion.}
Under these assumptions, we describe the prekey depletion attack on a user Bob, in which the PFS of initial messages sent to him is violated by constantly depleting the ephemeral (one-time) prekeys, usually automatically deposited by Bob on the WhatsApp servers. 
We assume Bob to be a security-cautious user with a very strict deletion policy regarding messages (i.e., disappearing messages set to the lowest value; currently 24h). Therefore, a compromise of his long- and medium-term security keys usually would not lead to a compromise of message content, if PFS guarantees hold, as the plaintext of messages would no longer be available on his device.  

In our attack, the attacker (Eve) with passive access to the end-to-end encrypted messages, tries to constantly deplete the ephemeral prekeys of a user Bob, s.t. a new session with Bob initiated by another user Alice will have no PFS for the initial messages sent from Alice to Bob. 
Note that it is not uncommon, even for long-term communication partners, to create new sessions with each other. This is mainly due to the increasing popularity of WhatsApp Web, which usually results in more short-lived browser sessions. An active already established, smartphone app session between both communication partners will not be directly affected by this attack though. 

Note that this paper focuses exclusively on point-to-point communication and does not consider group messaging scenarios.

\subsubsection{Privacy and Availability Attack Model}
\label{sec:privacy_threat_model}
In this case the attacker Eve is not interested in uncovering the content of Bobs conversations, but in gathering information about's Bob behavior and his devices and in denying that \emph{any} account can communicate with Bob via WhatsApp at all. Therefore, the goals in this case are as follows:

Although the objectives of the two attack models differ, both exploit the same underlying prekey mechanisms and functionalities. As a result, the attacks are closely related and interconnected, underscoring the inherent privacy and security challenges associated with this mechanism.

\begin{itemize}
     \item[\roundframe{G2}] \textbf{Device Status Tracking:} Monitoring the online/offline status and activity phases (active use vs. standby) of Bob's device(s).
     \item[\roundframe{G3}] \textbf{Fingerprinting:} Gather information about Bob's operating system (Android, iOS, macOS, Windows), their device's age and the number of (new) interactions within a given time span%
     .
     \item[\roundframe{G4}] \textbf{Denial of Service:} Deny any other account to establish a new communication session via WhatApp with Bob.
\end{itemize}

\parvspace
\noindent
\textbf{Attacker Capabilities.} For this attack model, the only requirements for the attacker are knowing the telephone number of the target and having a WhatsApp account.

\parvspace
\noindent
\textbf{Prekey Side-channel and Refills.} To achieve their goals, the attacker inspects subtile differences in the way the victim pushes fresh prekeys to the server.

\section{Background}
This section should provide a high level overview of the necessary protocol aspects of the Signal protocol. For more details we refer to Appendix~\ref{sec:signal}.

In this paper we target to the current WhatsApp adaption of the Signal protocol(s) described in their Whitepaper~\cite{noauthor_whatsapp_2023}, but since the official WhatsApp client software is not open source the exact implementation of the protocols is not easily obtainable.  
The origins of the Signal protocol, date back to the messaging App \emph{TextSecure}, started in 2010, which introduced a \emph{double ratchet} construction\footnote{Initially referred to as \emph{Axolotl Ratchet} to emphasize the self-healing properties of the protocol.} where communicating parties derive new keys for sending and receiving messages. 
The Signal protocol actually consists of an entire family of protocols~\cite{marlinspike_x3dh_2016, perrin_double_2016, marlinspike_private_2014, marlinspike_sesame_2017, kret_pqxdh_2023} which been studied in a variety of works~\cite{cohn-gordon_post-compromise_2016, alwen_double_2019, cohn-gordon_formal_2020, brendel_post-quantum_2022,wichelmann_help_2021, cremers_formal_2023, clone2020}.

In this paper we focus on the desired \emph{(perfect) forward secrecy} (PFS) guarantees of the handshake protocol and the practical implications regarding privacy to ensure it. NIST defines perfect forward secrecy, or just \emph{forward secrecy} for short, as follows: 

\emph{''Forward Secrecy (FS): Assurance obtained by one party in a key-agreement
transaction that the keying material derived during that transaction is secure against
the future compromise of the static private key-agreement keys (if any) of the
participants.''}~\cite{NIST800-56A-R3}

The Signal protocol uses three different types of Diffi-Hellman public keys to ensure forward secrecy right from the start: 
Long-term identity keys, medium-term (signed) prekeys and short-term (one-time) ephemeral prekeys (see Table~\ref{tab:keys_simple} for an overview). 
In our scenario Alice is the \emph{initiator} and wants to establish a secure connection with Bob (the \emph{responder}). 
The Signal protocol, as also implemented by WhatsApp, uses prekey bundles deposited by every user at a central server to allow any \emph{initiator} to negotiate a shared secret (via Diffi-Hellman Key exchange) even if the \emph{responder} is currently not online/available using the X3DH protocol~\cite{marlinspike_x3dh_2016}~\footnote{Since late 2023 Signal replaced X3DH by PQXDH~\cite{kret_pqxdh_2023}. On a high level, PQXDH is comparable to X3DH with the difference that an additional key from a post quantum key encapsulation mechanism (KEM) is used.}. 

\begin{figure}
    \centering
    \includegraphics[width=\linewidth]{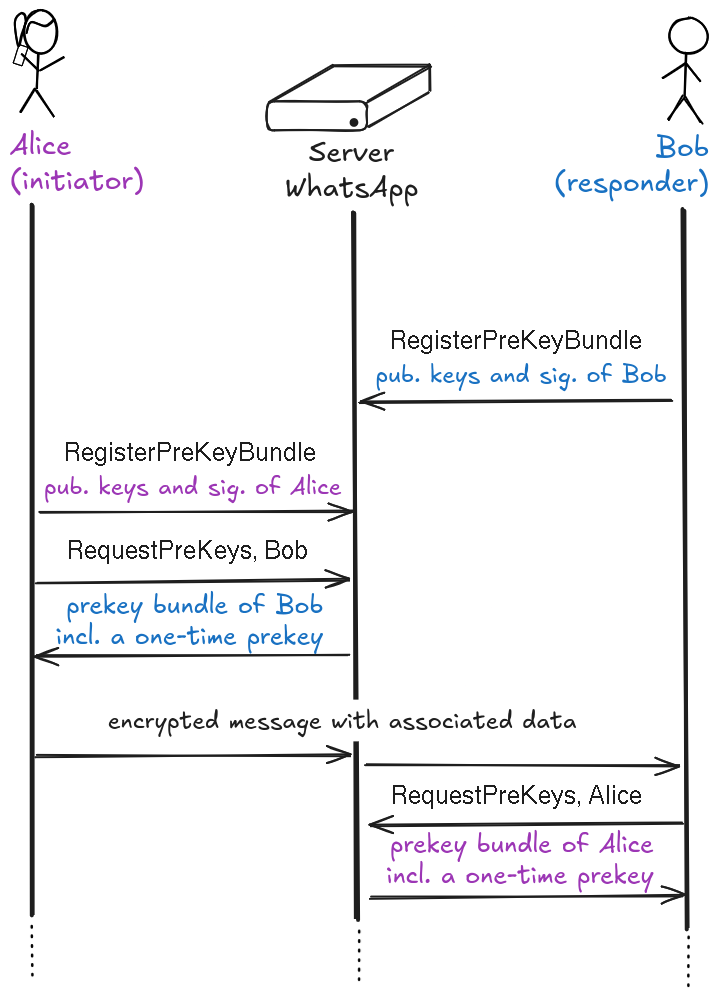}
    \caption{High-level overview of the intended prekey bundle data deposit and retrieval process, illustrating the ideal case in which everything functions as expected (i.e., prekey bundle of Bob contains also a one-time prekey). 
    }
    \label{fig:highlevel}
\end{figure}

\begin{table*}[]
\centering
\begin{tabular}{@{}lll@{}}
\toprule
Keys    &                 & Description                          \\ \midrule
$ipk^A$ & $ik^A $         & Long-term identity key pair of Alice \\
$ipk^B $ & \color{orange} $ik^B $      & Long-term identity key pair of Bob   \\
$prepk^A$ & $ prek^A  $  & Medium-term prekey pair of Alice, aka. \emph{signed prekey} \\ 
$prepk^B$ & \color{orange} $prek^B  $  & Medium-term prekey pair of Bob, aka. \emph{signed prekey} \\ 
$eprepk^A_n$ & $eprek^A_n $ & Short-term prekey pair number $n$ of Alice, aka. \emph{ephemeral prekey} or \emph{one-time prekey} \\
$eprepk^B_n$ & $eprek^B_n $ & Short-term prekey pair number $n$ of Bob, aka. \emph{ephemeral prekey} or \emph{one-time prekey} \\  \midrule
\multicolumn{2}{l}{$\langle ipk^A, prepk^A, Sig(ik^A, prepk^A), {\color{red}[ eprek^A_n ]}\rangle$} & A prekey bundle deposited by Alice on the server \\
\multicolumn{2}{l}{$\langle ipk^B, prepk^B, Sig(ik^B, prepk^B), {\color{red}[ eprek^B_n ]}\rangle$} & A prekey bundle deposited by Bob on the server \\
\bottomrule
\end{tabular}
\caption{Main cryptographic keys of the signal protocol relevant for our attacks. Public keys in asymmetric schemes always end in $pk$. The naming convention of the keying material is according to Cohn-Gordon et al.~\cite{cohn-gordon_formal_2020}. The ephemeral prekeys (one-time prekeys), which are considered optional in the prekey bundle, are depicted in {\color{red}red}. This is the key type drained in our attack. The secret keys of Bob which an attacker has to compromise to benefit from the violation of forward secrecy, i.e., if no ephemeral prekeys can be used, are depicted in {\color{orange}orange}.}
\label{tab:keys_simple}
\end{table*}

The initial handshake works as depicted in Figure~\ref{fig:highlevel} and starting with Formula~\ref{for:handshake_high_simple}. For more details we refer the reader to the Appendix~\ref{sec:signal}. First, the prekey bundle of the responder (in our case Bob) is fetched from the server by the initiator (in our case Alice). 
The information from the prekey bundle is verified by Alice through checking the signature on the \emph{signed prekey} using the (long-term) identity pubic key of Bob. As within other works~\cite{cohn-gordon_formal_2020}, it is assumed that Alice has already verified out-of-band that the long-term identity public key indeed belongs to Bob. 

Then the public keys from Bob's prekey bundle are used to compute shared keys for the ratcheting and message encryption and authentication. Here now we have to distinguish between the case where an ephemeral (one-time) prekey $eprepk^B$ of Bob is available or not. If no ephemeral prekey is available, the DH invocation $dh_4$ in Formula~\ref{for:dh4_simple} is omitted. Since this is the key type drained in our attack it is colored in {\color{red}red}. 

In any case, before initiating the session and sending the first message to Bob, Alice generates ephemeral keys including a ephemeral key pair $(ek^A, epk^A)$. Those are also used in the initial handshake and to initialize the DH ratchet construction, also referred to as the \emph{asymmetric ratchet}. 

\begin{align}
\label{for:handshake_high_simple}
dh_1  &\gets DH(ik^A,prepk^B) \\
dh_2  &\gets DH(ek^A,ipk^B)\\
dh_3  &\gets DH(ek^A,prepk^B)\\
\color{red} [dh_4 &\color{red} \gets  DH(ek^A,eprepk^B) ] \\
\label{for:dh4_simple}
rk_0 &\gets KDF_{r}(dh_1 \mid\mid dh_2 \mid\mid dh_3 {\color{red}[ \mid\mid DH4 ]})\\
\vdots \nonumber
\end{align}

At the end, a new message key $mk$ is derived using a key derivation function (KDF). This message key is then used to encrypt and authenticate ($AE$) the first chat \emph{message} from Alice to Bob, as well as to authenticate some associated data $AD$ consisting of the ephemeral public keys generated previously by Alice.
This is shown in Formula~\ref{for:init_simple} and~\ref{for:msg_simple}.

\begin{align}
\label{for:init_simple}
AD &\gets \langle \text{ephemeral public keys of Alice}, \\
   &\text{ id of } prepk^B, {\color{red} [ eprepk^B ] } \rangle \nonumber \\
\label{for:msg_simple}
AE_{mk}, AD &\gets E(mk,message,AD)
\end{align}

Once Bob receives this initial message, he can compute the same shared keys using his identity key ${\color{orange}ik^B}$ and his prekey ${\color{orange}prek^B}$, as well as the public keys of Alice consisting of her identity key $ipk^A$ (Formular~\ref{for:ipk_simple}) and her ephemeral handshake public keys (Formular~\ref{for:epk_simple} and~\ref{for:epk_simple2}).
Since the later are transmitted in the associated data $AD$, they are authenticated, but not encrypted. 
Tho benefit from the PFS violation (due to missing one-time prekys) and successfully decrypt recorded E2EE messages, the {\color{orange}orange} keys have to compromised by the attacker later on. 

\begin{align}
\label{for:ipk_simple}
dh_1 &= DH(ipk^A,{\color{orange}prek^B}) \\
\label{for:epk_simple}
dh_2 &= DH(epk^A,{\color{orange}ik^B})\\
\label{for:epk_simple2}
dh_3 &= DH(epk^A,{\color{orange}prek^B})\\
rk_0 &\gets KDF_{r}(dh_1 \mid\mid dh_2 \mid\mid dh_3)\\
\vdots \nonumber
\end{align}

The received message is then decrypted using the previously computed shared message key $mk$:
\begin{align}
    message,AD &\gets D(mk,AE_{mk}, AD)
\end{align}

If no ephemeral prekeys have been fetched by Alice, this initial message sent by Alice has no forward secrecy if observed by an attacker. 
Therefore, an attacker who is able to compromise Bobs medium-term and long-term secret keys $prek^B$ and $ik^B$ later on, can recompute the same message key $mk$, which highlights that there is no forward secrecy for this message. 
If Alice sends multiple messages before receiving any response from Bob, all these messages are affected as well, as the keys for these messages come from the symmetric ratchet. 
Forward secrecy is restored through the asymmetric ratchet, when Bob responds to a message. 
As soon as these ephemeral ratchet keys are deleted, forward secrecy is regained for this as well as subsequent messages.
Note, that even if forward secrecy is regained in a chat session, the initial messages sent from Alice to Bob have been encrypted using the symmetric ratchet only.
Therefore, they have no forward secrecy for the entire lifetime of the signed prekey $prek^B$ of Bob.
According to the specifications, the signed prekey should be periodically rotated~\cite{noauthor_whatsapp_2023,marlinspike_x3dh_2016,meta2023messenger}, where suggested intervals reach from once a week to once a month~\footnote{In practice Signal rotates the signed prekey every two days~\url{https://github.com/signalapp/Signal-Android/blob/481dc162d80292a046b4229cceba2ac2f2a73f36/app/src/main/java/org/thoughtcrime/securesms/jobs/PreKeysSyncJob.kt\#L57-L66}}. WhatsApp refreshes signed prekeys usually once every month. To the best of our knowledge, the open source implementations whatsmeow, baileys and cobalt never replace the initially uploaded signed prekey, which of course would increase the impact of a loss in forward secrecy.

\section{Testing Environment}

To effectively test WhatsApp's session and encryption procedures, we set up a test and experimentation environment which we describe in this section.
To capture and understand low-level protocol messages, we base our work on existing community projects (i.e., unofficial WhatsApp clients that are based on reverse-engineering of the official implementation).

\begin{table*}[]
\centering
\begin{tabular}{@{}lrrl@{}}
\toprule
Project        & GitHub Stars & Lines of Code & Project Scope                                                     \\ \midrule
Baileys        & 4,856        & 134,052 & Emulating WhatsApp Web (companion) devices                        \\
whatsmeow      & 2,549        & 67,088 & Emulating WhatsApp Web (companion) devices                        \\
Cobalt         & 708          & 41,115 & Emulating main (Android/iOS) and WhatsApp Web (companion) devices \\
CobaltAnalyzer & 37           & 331 & Capturing decrypted traffic of legitimate WhatsApp Web browser sessions     \\ \bottomrule
\end{tabular}
\caption{Relevant WhatsApp community projects and their offered features that we used throughout our analysis.}
\label{tab:community-projects}
\end{table*}

We've used the community projects that are shown in Table~\ref{tab:community-projects} to dynamically send and inspect requests from our own WhatsApp devices (Android, iOS, Web, Desktop). Furthermore, we wrote a custom client that allows querying the existing WhatsApp devices and their cryptographic keys ($ipk$, $prepk$, $eprepk$) for an arbitrary telephone number.

\subsection{Relevant Endpoints and Message Structs}
To uniquely identify and address users within the messaging service, WhatsApp uses so-called JIDs (\textit{Jabber ID}), following a specific addressing scheme: \mbox{\texttt{<phoneNumber>@<serverName>}}.

Every device that is registered for a specific phone number, gets their own \emph{device ID}.
The device ID is an auto-incrementing index for each user, that is reset whenever the user sets up WhatsApp on their phone.
Device 0 always represents the main device (i.e., the smartphone), while non-zero device IDs are used for companion devices (i.e., web- or desktop clients).
To address specific devices within a JID, the device ID is encoded between the phone number and the server name: \mbox{\texttt{<phoneNumber>:<deviceId>@<serverName>}}.
For example, \mbox{\texttt{123456789:1@s.whatsapp.net}} represents the first companion device that is registered for the US-based phone number \mbox{\texttt{+123456789}}.

Using our custom client, we can fetch the available device IDs for an arbitrary phone number:
\begin{myTerminal}
./query-devices -t 123456789`\StartTerminalConsole`
Querying registered devices for target number.

Found 3 existing devices: [0, 1, 3]`
\end{myTerminal}

In this case, the target has one main device (index 0) and two companion devices (index 1 and 3).
Since no device with index 2 is available in this list, we can deduce that there has been a linked device (e.g., a desktop computer or WhatsApp web session) with index 2, which has been logged out (unlinked). At some later point a new device has been linked, which due to the auto-incrementing nature of the device IDs, now has index 3. 

WhatsApp's encryption scheme requires the message sender to individually encrypt and send messages for every device of the recipient.
Thus, the sender can query a users' current device list from the server via so-called \texttt{usync infoqueries}.
WhatsApp also allows sending messages to new contacts (or unknown phone numbers), thus this endpoint is not only limited to known contacts, but can also be queried for external numbers.
Using our testing client, we can retrieve (and consistently monitor) the registered devices and their corresponding device IDs for arbitrary phone numbers as already demonstrated in~\cite{gegenhuber_2024_carelesswhisper}.

Besides knowing the recipient's device list, the sender also needs to retrieve each device's DH keys, to individually encrypt the message for every target device.
Again, the corresponding \texttt{inforquery} can be issued to retrieve a \emph{prekey bundle} for an arbitrary phone number.
Listing~\ref{lst:prekey_response} shows an example for the information that is returned by this query.
In summary, the endpoint returns, 
\begin{itemize}
    \item the three DH public keys ($ipk$, $prepk$, $eprepk$).
    \item their corresponding \emph{key IDs} (registration ID, signed prekey ID and one-time prekey ID).
    \item the signature of the signed prekey $prepk$.
    \item an epoch timestamp indicating when the device last updated the relevant object (i.e., pushed a new $prepk$ or $eprepk$ to the server).
\end{itemize}

\footnotetext{
Baileys: %
\href{https://github.com/WhiskeySockets/Baileys}{\texttt{github.com/WhiskeySockets/Baileys}}\\
whatsmeow: %
\href{https://github.com/tulir/whatsmeow}{\texttt{github.com/tulir/whatsmeow}}\\
Cobalt: %
\href{https://github.com/Auties00/Cobalt}{\texttt{github.com/Auties00/Cobalt}}\\
CobaltAnalyzer: %
\href{https://github.com/Auties00/CobaltAnalyzer}{\texttt{github.com/Auties00/CobaltAnalyzer}}%
}

The \emph{key ID} of the signed prekeys and the \emph{key ID} of the one-time prekeys are also incremented with every new key (of the respective type) uploaded to the server, but these IDs do not always start with zero (for more information we refer to Section~\ref{sec:fingerprinting}).

While the device's long-term (static) identity key and the medium-term signed prekey typically remain unchanged across subsequent queries, the included one-time prekey changes with each query. 
As the name suggests, this prekey is intended for single use and is therefore discarded from the server after being disclosed to a third party.
In practice, the endpoint is not called very often from a single account, as a device's prekey bundle only needs to be retrieved from the server when initiating a new session.
For active sessions, the key material is continuously renewed and embedded within ongoing direct messages between the communicating parties.
However, as we show, a malicious actor can deliberately request multiple prekey bundles from the server, effectively draining a device's one-time prekey reserve that is saved at the server.

\begin{lstlisting}[caption={WhatsApp prekey bundle containing identity key, signed prekey and a single one-time prekey, queriably for arbitrary phone numbers. The values for the byte arrays are shortened due to the limited space.}, label=lst:prekey_response, float, language=json]
{
  "jid": "123456789:1@s.whatsapp.net",
  "t": "1740182155"       // epoch timestamp
  "registration": "000005DB",
  "type": "05",           // key type (djb)
  "identity": "76..77",   // 32 bytes pubkey
  "skey": {               
    "id": "000001",       // signed prekey
    "value": "44...6a",   // 32 bytes pubkey
    "signature": "0d..02" // 64 bytes
  },
  "key": {                
    "id": "0001a",        // one-time prekey
    "value": "0e..0b"     // 32 bytes pubkey
  }
}
\end{lstlisting}

\subsection{Testing Methodology}
To assess whether an attacker can deplete the saved prekeys of a specific target device and to compare official implementations and the impact across different device categories, we systematically retrieve and analyze the returned key values in various settings and scenarios.

\subsubsection{Server-side Prekey Output and Rate Limits} %
In our custom client, we've implemented a function that queries the prekeys for a specific target device repeatedly.
By targeting our own devices, we want to investigate whether there are server-side protections or rate-limiting mechanisms, stopping an attacker from doing so.
Furthermore, we want to test how fast an attacker is able to consume prekeys and whether consistent prekey depletion is potentially feasible.
Finally, we check whether the server properly removes the prekeys, or whether certain keys are accidentally returned more than once.

\subsubsection{Client-side Prekey Input and Push Behavior}
According to the protocol design, client devices push new prekeys to the server whenever necessary.
We want to investigate how many prekeys are typically saved on the server and under which circumstances they're refilled.

\parvspace
\noindent
\textbf{Prekey Reserve Batches.}
Our analysis covers different device types (Android, iOS, Windows, macOS, Web) and different prekey states (initial reserve vs. prekey refills).
To measure the amount of prekeys that were pushed in different client states, we consume (and count) all available prekeys for our target device using our custom client.
To remove additional noise during the measurement and to prevent the target device from immediately refilling the prekey buffer, we put it into flight mode.

\parvspace
\noindent
\textbf{Prekey Refill Trigger.}
After measuring typical prekey batch sizes, we repeat the above experiment without putting the target device into flight mode.
A prekey refill also updates the epoch timestamp that is sent alongside the prekey bundle \texttt{infoquery} response (cf. Listing~\ref{lst:prekey_response}).
Thereby, we can also observe when the last prekey refill mechanism was triggered for the target device.

\parvspace
\noindent
\textbf{Fingerprinting Device Types and Activity States.}
Besides identifying general conditions that trigger a prekey refill, we want to investigate whether different device types or activity states (e.g., standby) influence how fast the prekey reserve is refilled.

\subsubsection{Exploration and Exploitation}
After understanding the design decisions of the official WhatsApp clients and potential rate-limits on the server, we outline various exploitation scenarios.
To test and verify them in practice, we use our custom client to target our own testing devices.
A detailed list of the used devices can be found in Section~\ref{tab:testing-devices} in the Appendix.

\section{Results and Exploitation}

Our tests showed that depleting a target's prekeys is possible and that WhatsApp currently employs little to no countermeasures against it. 
We furthermore show our detailed results and outline the abuse potential and specific security- and privacy implications.

\begin{table}[t]
\centering
\begin{tabular}{@{}ll|ll|ll@{}}
                          & \multicolumn{1}{l|}{}           & \multicolumn{2}{c|}{Standby}                             & \multicolumn{2}{c}{Screen On}                           \\
                          & \multicolumn{1}{l|}{Device}     & \multicolumn{1}{c}{WiFi}   & \multicolumn{1}{c|}{4G}     & \multicolumn{1}{c}{WiFi}   & \multicolumn{1}{c}{4G}     \\ \midrule
\faApple   & \multicolumn{1}{l|}{iPhone SE}  & \multicolumn{1}{r}{85\%} & \multicolumn{1}{r|}{94\%} & \multicolumn{1}{r}{90\%} & \multicolumn{1}{r}{93\%} \\
\faApple   & \multicolumn{1}{l|}{iPhone 8}   & \multicolumn{1}{r}{90\%} & \multicolumn{1}{r|}{88\%} & \multicolumn{1}{r}{89\%} & \multicolumn{1}{r}{88\%} \\
\faApple   & \multicolumn{1}{l|}{iPhone 11}  & \multicolumn{1}{r}{80\%} & \multicolumn{1}{r|}{96\%} & \multicolumn{1}{r}{74\%} & \multicolumn{1}{r}{80\%} \\

\faAndroid & \multicolumn{1}{l|}{Poco X3}    & \multicolumn{1}{r}{76\%} & \multicolumn{1}{r|}{55\%} & \multicolumn{1}{r}{18\%} & \multicolumn{1}{r}{17\%}   \\
\faAndroid & \multicolumn{1}{l|}{Galaxy A54} & \multicolumn{1}{r}{10\%} & \multicolumn{1}{r|}{9\%}  & \multicolumn{1}{r}{18\%} & \multicolumn{1}{r}{4\%}  \\
\faAndroid & \multicolumn{1}{l|}{Redmi 10}   & \multicolumn{1}{r}{15\%} & \multicolumn{1}{r|}{72\%} & \multicolumn{1}{r}{13\%} & \multicolumn{1}{r}{19\%}   \\
\end{tabular}
\caption{Success rate among different devices, i.e., how likely it is that a prekey bundle fetched for a new session will \emph{not} contain a one-time prekey even if the target is currenlty online and powered on and an attacker is actively depleting its one-time prekeys.}
\label{tab:device-success-rates}
\end{table}

\subsection{Perfect Forward Secrecy Degradation}
\label{sec:result-pfs-attack}
For this section, we consider the \emph{PFS attack model} described in Section~\ref{sec:pfs_threat_model}, so the goal \roundframe{G1} of the attack is it to degrade PFS for new sessions initiated with Bob by other users (such as Alice). Therefore, Eve wants to deplete all one-time prekeys of Bob. 

\begin{figure}
    \centering
    \includegraphics[width=\linewidth]{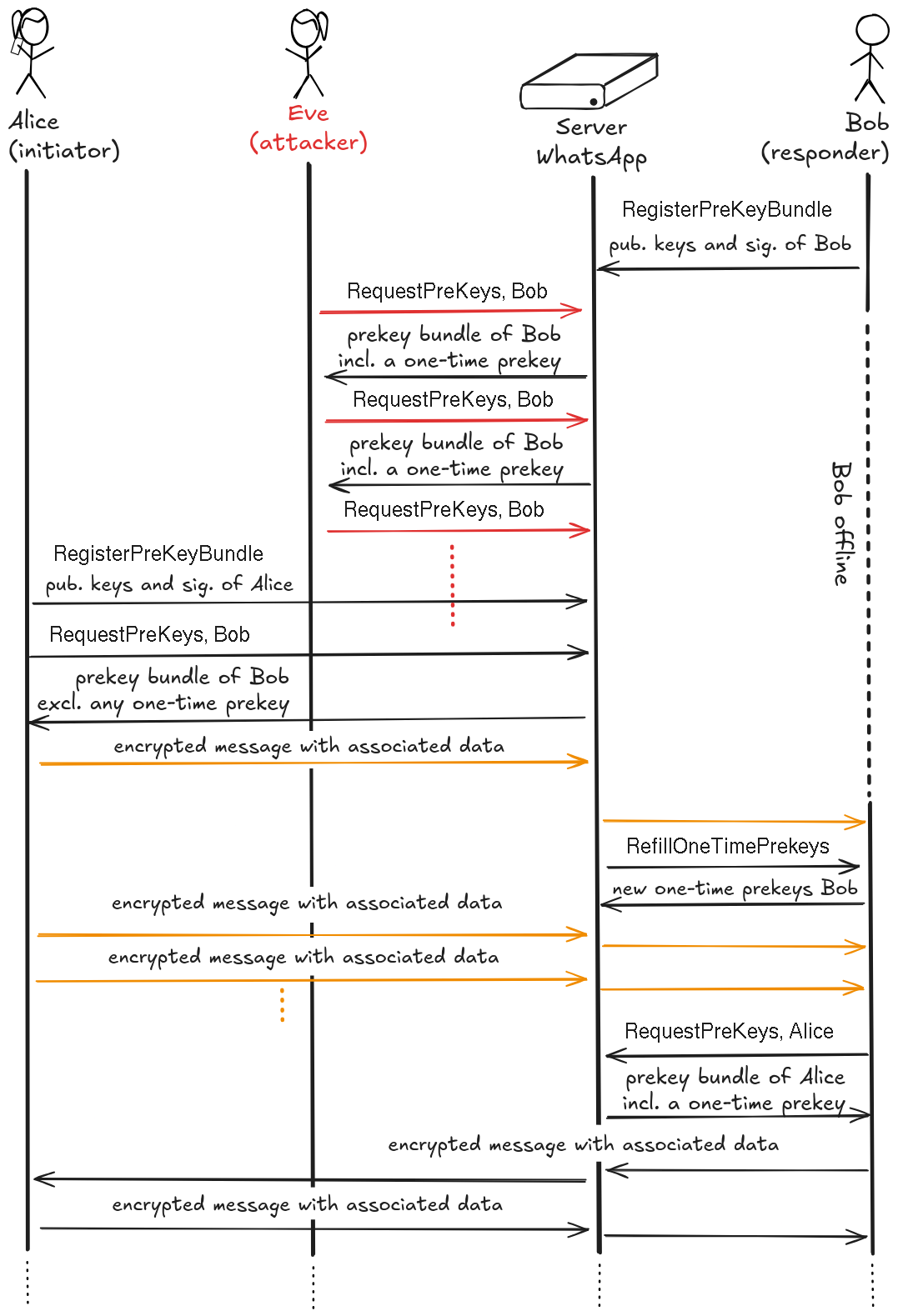}
    \caption{High-level overview of the attack. Eve is flooding the server with requests for prekey bundles of Bob. This prevents any other user, such as Alice, from obtaining a one-time prekey of Bob. Therefore, in this case all messages from Alice to Bob ({\color{orange}orange}) do not have forward secrecy, as long as the associated secret key of the signed prekey belonging to Bob is not deleted. New messages in this session after a response of Bob are not affected since new ephemeral keys are used and therefore forward secrecy is regained as soon as these new ephemeral keys are deleted.}
    \label{fig:highlevelattack}
\end{figure}

For our first attack we consider the simplest case, where the targeted device is currently offline and thus cannot refill depleted one-time prekeys at the moment. This attack is depicted in Figure~\ref{fig:highlevelattack}. 
Our tests showed, since WhatsApp does not enforce any rate limiting, constantly querying prekey bundles for any user is possible, thereby eventually depleting all available one-time prekeys. 
Although one-time prekeys are crucial for ensuring PFS, messages can also be transmitted when only identity and signed prekey are available to generate a shared secret (i.e., when no one-time prekeys are available on the server).
In our tests, this PFS degradation is \emph{not} indicated to the users UI (e.g. by a security notification in the WhatsApp client initiating the session). Therefore, the PFS degradation attack can be executed without the affected users noticing in their UI. 
Note that even if Bob's device comes online after Alice has initiated a new communication session without a one-time prekey, messages sent by Alice will not achieve PFS until Bob's device responds for the first time within that session.

Now let's consider the case where the targeted device of Bob is constantly online. In this case, it receives a notification from the server as soon as there are less than 11 one-time prekeys left. Depending on the device type and its current power saving state, it reacts upon this notification and uploads 812 new one-time prekeys to the server. 
For measurements on how fast depletion of one-time prekeys is possible in case the targeted device is online, as well as the detailed behavior and availability implications, see Section~\ref{sec:characteristic-refills} and~\ref{sec:rapid}. 
Table~\ref{tab:device-success-rates} provides an overview of the measured success probabilities of the one-time prekey depletion attack in case the targeted device is online. In other words, it indicates the likelihood that another party (e.g., Alice) will initiate a new communication session with Bob without using a one-time prekey while our depletion attack is in progress.

\subsection{Device Online Status Leak}
For this and the remaining section we consider the \emph{privacy and availability attack model} described in Section~\ref{sec:privacy_threat_model}. In this section we focus on the goal \roundframe{G2} \emph{device status tracking}.

\begin{figure}[tb]
    \centering
    \includegraphics[width=\linewidth]{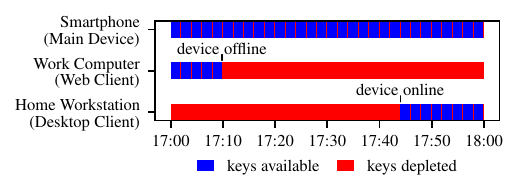}
    \caption{Each device's online status can be independently, consistently and stealthily monitored, possibly leaking the victim's location and daily routines.}
    \label{fig:companion-online-status-monitoring}
\end{figure}

A device can only push fresh prekeys to the server, if it is turned on and connected to the Internet.
When a device's prekeys are drained by an attacker and the reserve on the server drops to less than 11 prekeys, the targeted device will receive a notification from the server to refill its one-time prekeys.
Consequently, depending on whether or not the one-time prekeys are refilled timely, this could also leaks the current online state of this particular device: If the one-time prekeys are not refilled timely, the targeted device is probably offline.
Furthermore, any refill will update the corresponding epoch timestamp of the prekey bundle (cf. Listing~\ref{lst:prekey_response}), which thereby can be seen as a lower bound for the last online time of the respective device.
Since this attack can be executed independently for every device of the victim, it can be used for stealthy and consistent tracking of connection states throughout the day.
This is especially critical for companion devices that are usually not always online, such as desktop computers.
For example, Eve could use this to track the victim's daily routines (and thus, corresponding locations) when switching between devices, as shown in Table~\ref{fig:companion-online-status-monitoring}.
To match devices to a specific context or location (e.g., work vs. home), the attacker could monitor the devices over a period of multiple weeks.

Main devices are usually always online and are thus less prone to this attack.
Nevertheless, omission of prekey refills for extended periods could still leak information about their current activity.
For example, the attack could be used to determine whether a person is currently on a flight, in a shielded building, or other locations without any Internet connection.
Delayed refills could hint to blind spots in cellular reception, e.g., when driving through a tunnel.
Finally, some people use the phone's flight mode to mute all notifications during the night, disclosing a person's sleep schedule.
The next Section addresses the question, how one-time prekeys are actually refilled by different devices and what information can be gathered through this feature. 

\begin{table*}[t]
\centering
\newcommand\tFa{$^{\mathrm{a}}$}
\newcommand\tFb{$^{\mathrm{b}}$}
\begin{tabular}{@{}l|ccc|cc|c@{}} \toprule
 & \multicolumn{3}{c|}{Initialization Values for \emph{key IDs} }& \multicolumn{2}{c|}{Prekey Batch Size} & \\ 
Client Implementation & Registration  & Signed PK   & One-Time PK  & Initial   & Refill      & Refill Trigger \\ \midrule
Android               & $R$           & 0           & $R$          & 812       & 812         & 10             \\
iPhone                & $R$           & $R$         & 1            & 812       & 812         & 10             \\ \midrule
WhatsApp Web\tFa      & $R$ \& 0x3FFF & 1           & 1            & 200       & 812         & 10             \\
Desktop App macOS     & $R$           & $R$         & 1            & 200       & 812         & 10             \\
Desktop App Windows   & $R$           & 1           & 1            & 50        & 812         & 10             \\ \bottomrule
\end{tabular}
\\
\vspace{1ex}
$R$~Random number. \hspace{1ex} %
\tFa~Verified on Firefox, Chrome, Safari. 
\caption{Different initialization values, prekey batch sizes, and incrementing ID patterns used across various implementations enable device fingerprinting (e.g., OS, device age). Beyond being a privacy risk, this could be exploited by an attacker during the reconnaissance phase to tailor further attacks.}
\label{tab:device-fingerprinting}
\end{table*}

\subsection{Device Fingerprinting}
\label{sec:fingerprinting}
Table~\ref{tab:device-fingerprinting} shows characteristic \emph{key ID} initialization values and particular prekey batch sizes used when uploading fresh keys to the server.
Different client implementations use different initialization values for assigning initial key IDs.
While this also holds true regarding the initial prekey batch size, all implementations use a fixed refill batch of 812 elements.
Furthermore, we did not see any difference regarding the refill trigger, as all clients push a new prekey batch (812 elements) to the server as soon as it has only 10 remaining prekeys left.
In detail, this is triggered by a server-side notification that is received by the client.
Pushing new one-time prekeys to the server always invalidates any previous one-time prekeys
After the initial assignment, signed prekeys and one-time prekeys are issued by incrementing their IDs.

\parvspace
\noindent
\textbf{OS Disclosure.}
Finding out a target's operating system might be a valuable information for an attacker during the reconnaissance phase to efficiently prepare for further attacks.
Due to different strategies for the used initialization values and the fact that we can naturally differentiate main and companion devices by their \emph{device IDs} ($0$ for main devices, $> 0$ for companion devices), we can distinguish the used client implementation and runtime environment for an arbitrary device with high confidence.
For users utilizing WhatsApp web, all operating systems and browsers showed the same behavior (because all browsers execute the same javascript code).

Fingerprinting a target primary device by characteristic \emph{signed prekey IDs} (e.g., Android vs. iOS) is possible with high confidence and just requires the attacker to query the prekey bundle once for the victims phone number and device ID $0$.
If the the signed prekey ID is high ($ >> 0 $), than it has probably be chosen at random and thus it must be an iPhone. 

Distinguishing companion device types by their initial prekey batch sizes requires repeated querying, but is still  feasible.
Due to homogeneous refill batch sizes across all device types, the initial prekey batch size can be deduced at any later point in time, by consuming the entire (812 element-sized) prekey batch from the server and looking at the returned minimum and maximum prekey IDs.

To facilitate this kind of OS fingerprinting, our client collects all fetched prekey bundles and summarizes the stats of the returned values:

\terminalSU{user=pogo,host=prekey,color=lime}
\begin{myTerminal}
./deplete-keys -t 123456789`\StartTerminalConsole`
Fetching all available prekeys of target device...

All prekeys depleted, consumed 812 bundles in 45s.
[Prekey Stats] Cnt: 812, MinID: 1674, MaxID: 2486`
\end{myTerminal}

In the above example, we know that the victim uses a Windows desktop application, where a batch of 50 prekeys were uploaded in the initial batch (corresponding calculation: $1,674 - 812 - 812 = 50$).
If the initial batch size would be $200$, we could still distinguish between macOS and WhatsApp Web by the value of the signed prekey ID as described before. 

\parvspace
\noindent
\textbf{Device Age and Activity Score.} %
The registration ID and identity key are static for the lifetime of the device installation.
In contrast, signed- and one-time prekeys (and their IDs) change over time.
When the device updates the signed prekey (usually done approx. once a month in WhatsApp) it also increments the corresponding signed prekey ID.
Thus, for all devices except iOS and macOS, the current signed prekey ID roughly corresponds to the device's age in months.

Similarly, one-time prekey IDs are incremented by all devices.
When asked by a client, the server always randomly selects one of the available one-time prekeys, making it more cumbersome to monitor for the attacker.
However, when consuming all available prekeys, the attacker can still deduce how many prekeys have been used since the last refill, by calculating the difference between 812 and the number of returned prekeys.
For all devices that initialize the one-time prekey with 1 (all except Android), the attacker can also derive the total amount of used one-time prekeys since the initial setup of the device.
Due to the natural depletion of the one-time prekeys that is caused by new people (or devices) contacting the victim, this can be used to estimate an activity- or chattiness score of the target.

\subsection{Observing Characteristic Refill Behavior}
\label{sec:characteristic-refills}
While measuring the approximate success rates for our PFS downgrade attack (Table~\ref{tab:device-success-rates} in Section~\ref{sec:result-pfs-attack}), we noticed that the refill behavior can vary widely, depending on the victim's phone and its current state (e.g., current access technology and screen on/off state).

For example, across all captured experiments, iPhones were more vulnerable to prekey depletion --thus, took longer to react on a drained prekey bundle-- than Android devices.
Among the various Android models, the Samsung Galaxy A54 consistently showed the fastest refill times. %

In addition to identifying the victim's operating system through specific key ID values, as presented in Section~\ref{sec:fingerprinting}, continuous monitoring of refill behavior could help determine the victim's operating system or device model.

Interestingly, the Xiaomi Poco X3 also showed significantly faster reactions when the screen was active compared to when it was in standby mode.
This aligns with previous work~\cite{gegenhuber_2024_carelesswhisper}, exposing activity fingerprinting by measuring message RTT times via delivery receipts.

Characteristic examples for the monitored refill behavior of different phones can be found in Figure~\ref{fig:characteristic-device-behaviour} in the Appendix.

\subsection{Rapid Retrieval and Denial of Service}
\label{sec:rapid}
For the previously presented attacks, our client sends synchronous queries to the server.
Thus, before sending another query requesting the victim's prekey bundle, we always waited for the response of the previous request, leading to a depletion rate of up to 20 prekeys per second.
In practice, the time for every request is limited by the connection round trip time (RTT) between client and server.
Furthermore, the process seems to be significantly influenced by the current server load, with the depletion of all 812 prekeys taking anywhere from 40 seconds to 2 minutes within our different experiments.

One-time prekeys are supposed to be returned just once, which requires synchronization across concurrent requests.
To test for the maximum retrieval rate for a single session%
, we increased the request rate, by sending queries in an asynchronous manner.
Using parallel requests, we were able to consistently deplete 812 prekeys within just 10 seconds.
After a certain retrieval rate (roughly more than 50 requests per second), the server occasionally returns a \texttt{503 Service Unavailable} instead of the actual prekey bundle.

We observed that the \texttt{503} server error is not merely a rate limit affecting the current client session.
Instead, generating a high volume of requests in one session also causes unsuccessful queries for other unrelated clients requesting prekeys for the same device for all querying devices.

By further increasing the request rate to > 2,000 requests per second, we can entirely clog the prekey retrieval for the corresponding victim device.
We showed the feasibility of this attack by clogging prekey retrieval with one session and concurrently trying to retrieve a valid prekey bundle by an unrelated client.

\parvspace
\noindent
\textbf{Denial of Service.}
In Section~\ref{sec:result-pfs-attack}, the attacker only tries to eliminate the one-time prekey layer from the key exchange, thus, starting a new conversation is still possible.
In contrast, the failure to retrieve the entire prekey bundle will generally hinder any new conversation attempts with the victim.

We verified this in practice by trying to contact our victim from different phones while simultaneously clogging prekey bundle retrieval using our custom client implementation.
In all cases, the phones were not able to effectively start a new session and send the corresponding message, but kept stuck at the \mbox{sending symbol~\clockMessageSending}).
Also, trying to audio/video call the target via WhatsApp resulted in the call being immediately dropped\footnote{According to a technical report of Meta, the call initiation should have been unaffected, as  the required prekey bundle is sent directly via webRTC~\cite{meta2023messenger}, as the communication partner needs to be online anyway to establish a call.}.
To demonstrate the attack in practice, we prove a short demonstration video\footnote{\url{https://drive.proton.me/urls/H2P7VCW9R4\#6ZGjnwFjf4TT}}.

We tried to infer the retransmission strategy empirically.
Thereby, we did not see any timing-based back-off strategies for message retransmission (i.e., even when we stopped our DoS attack, the corresponding clients were not automatically trying to resend the message).
Re-entering the chat or minimizing/activating the application to/from standby does not automatically trigger a retry-procedure.
However, when entirely closing and reopening the WhatsApp application, the client makes another attempt to request the prekey bundle and eventually transmits the message to the target.
Our testing attempts to execute this attack over extended time periods (e.g., multiple hours) were not blocked by WhatsApp, thus performing this attack consistently seems currently feasible.
Thereby, an attacker could completely prevent any new conversation attempt to the target and thus force a downgrade to use less secure messaging solutions (e.g., SMS, Telegram).

\subsection{Battery Drainage}
Consistently generating new prekeys and pushing them to the server results in additional battery drain and likely prevents the phone from entering deep sleep states.
We measured this extra drain under conditions of rapid (i.e., asynchronous) prekey depletion, which forces the device to frequently regenerate and upload fresh prekeys.
For this case, we use our Samsung Galaxy~A54~5G as target, since it refilled prekey almost immediately, even when being put into standby mode (cf. Section~\ref{sec:characteristic-refills}), presumably increasing abuse potential and battery drain.
Our measurements showed an additional battery drain of approximately 2\% per hour (measured during standby in LTE).
During this time, prekeys were depleted roughly every 15 seconds, and the process of uploading new prekeys resulted in about 8~MB of additional data usage per hour.
While this attack could definitely be annoying for the victim (i.e., forcing them to recharge their phone throughout the day), we consider this only a minor availability issue.
Nevertheless, similar to the previously discussed vulnerabilities, any WhatsApp user can be targeted covertly, with minimal evidence left behind on the victim’s device.

\subsection{Peripheral Observations}
Besides the presented exploits, we made additional observations that could be relevant for WhatsApps security or privacy.

\parvspace
\noindent
\textbf{Desynchronization of Prekey Depletion State.}
During rapid prekey depletion, we observed instances where the prekey state between the server and client became desynchronized. 
As a result, the client was not aware that the prekeys had been drained and thus did not push fresh prekeys to the server.
In practice, this increases the success rate for our PFS depletion attack.

Additionally, when inspecting the decrypted traffic of legitimate WhatsApp web clients (using \textit{CobaltAnalyzer}), we observed that prekey refills of the client were occasionally rejected with a \texttt{503 Service Unavailable} error.
While the client was eventually able to upload a fresh prekey bundle, this of course also enlarges the available time window for a PFS degradation attack.

\parvspace
\noindent
\textbf{Repeated Prekey Distribution.}
In the course of depleting prekeys from our test devices to evaluate the effectiveness of the PFS downgrade attack, we collected and analyzed the returned prekey bundles to verify whether one-time prekeys were correctly discarded after use.

While the server generally behaved as expected --successfully synchronizing concurrent queries from two independent sessions targeting the same device-- we did observe rare instances in which one-time prekeys were handed out more than once.
In total, we documented four such occurrences of prekey reuse, potentially indicating isolated failures in the server's prekey distribution.

\parvspace
\noindent
\textbf{Omitted Prekey IDs (Android).}
While adhering to the uniform batch-size of 812 elements, we noticed that for every additional prekey batch that is uploaded to the server, 2 prekey IDs are omitted by the Android client.
This behavior suggests the presence of a potential off-by-two error in the Android client's prekey generation implementation.
While simply skipping prekey IDs is not a security issue by itself, it could again by abused to determine the victim's operating systems as presented in Section~\ref{sec:fingerprinting}.

\parvspace
\noindent
\textbf{Blocked Contacts.}
Because the attacker interacts only with the server and never sends direct packets to the victim, the victim has no means of identifying the source of the attack.
Nevertheless, we evaluated whether blocking an account in WhatsApp has any impact by testing a scenario in which the victim had already blocked the attacker prior to the attack.
The results show that blocking has no effect, thus the attack remains fully feasible.

\section{Related Work}

\parvspace
\noindent
\textbf{Security and Privacy at Instant Messaging.}
Schrittwieser et al. were the first to investigate security and privacy issues in mobile instant messaging and VoIP applications, uncovering vulnerabilities such as account hijacking and number spoofing~\cite{schrittwieser_2012_guess, mueller_2014_what}.
Beyond Over-the-Top (OTT) applications, similar security vulnerabilities have been identified in VoIP-based messaging solutions like VoLTE, VoWiFi and RCS~\cite{tu_2016_new, gegenhuber_2023_mobileatlas, gegenhuber_2024_diffie, yang_2024_uncovering}.
In many cases, these vulnerabilities remained undiscovered for years due to proprietary clients and the lack of tooling for security experiments.
We adopt a similar security testing approach to find vulnerabilities within WhatsApp, leveraging open-source tools to emulate a real client sending protocol queries to the server.
In contrast, Hagen et al.~\cite{hagen_2021_all} demonstrated that large-scale account enumeration is possible in major messaging applications (e.g., WhatsApp, Signal) simply by automating interactions with the regular user interface.
Beyond that, efficient account enumeration can also be achieved by using open-source clients to directly extract data from internal user APIs~\cite{gegenhuber_2025_heythere}.

\parvspace
\noindent
\textbf{Fingerprinting and Side Channels.}
Previous work has shown that convenience features, such as read receipts in WhatsApp and other instant messaging apps, are frequently misused for stalking, even by non-technical users~\cite{freed_stalker_2018}.
Beyond read receipts, delivery receipts expose message round-trip times (RTTs), which can be exploited to infer a user's coarse geolocation~\cite{schnitzler_hope_2023}.
Recent work~\cite{beery_whatsapp_2024} has further revealed that WhatsApp’s multi-device feature inadvertently leaks users' device lists and the specific device used to send a message, due to the design approach used for end-to-end encryption (E2EE).
Moreover, Gegenhuber et al. have demonstrated that read receipts can be leveraged for malformed and thus invisible messages in multi-device settings, enabling independent tracking and fingerprinting of all a user’s devices, as well as their online status and operating system~\cite{gegenhuber_2024_carelesswhisper}.
As a potential mitigation, WhatsApp now allows users to block messages from unknown accounts\footnote{\url{https://faq.whatsapp.com/3379690015658337/}}.
However, while our prekey depletion exploit can be used for similar fingerprinting techniques --such as tracking a user's device online status-- we do not send any direct messages to the victim’s phone.
Instead, our approach interacts solely with WhatsApp's central prekey server.

\parvspace
\noindent
\textbf{Signal Protocol.} The Signal Protocol and its variants has been analyzed in a series of works~\cite{cohn-gordon_formal_2020, cohn-gordon_post-compromise_2016, chase_signal_2020, brendel_post-quantum_2022, wichelmann_help_2021, cremers_formal_2023, clone2020}. 
Cohn-Gorden et. al~\cite{cohn-gordon_formal_2020} provides a nice overview of the protocol as well as a  formal security analysis of the triple Diffie-Hellman (X3DH) key agreement and the Double Ratchet (DR). 
The DR was initially analyzed in~\cite{alwen_double_2019}. A new variant of the key agreement, called PQXDH, which includes PQ-KEM as been described and formally analyzed in~\cite{kret_pqxdh_2023}.
Post-Compromise Security (PCS) was initially defined and analyzed in~\cite{cohn-gordon_post-compromise_2016}. Attacks on PCS in a multi device setting have been described in~\cite{wichelmann_help_2021, clone2020}. 
The entire conversation layer, potentially consisting of multiple sessions/devices of a user, has been analyzed in~\cite{cremers_formal_2023} also with a focus on PCS and cloned devices.

\section{Discussion}

\subsection{Ethical Considerations}
For our measurements and during experimentation we only targeted WhatsApp accounts under our direct control.
Additionally, we tried to adhere to WhatsApp's protocol through the use of community-proven open source projects (some of them being used in widely deployed production systems~\footnote{\url{https://github.com/element-hq/mautrix-whatsapp}}).
In our depletion experiments, we issued a substantially higher volume of prekey queries compared to typical client implementations.
However, this increased traffic is unlikely to pose a significant risk to WhatsApp’s infrastructure, which is built to serve more than three billion users.
Moreover, we limited our offensive depletion to at most two concurrent client sessions.
Lastly, none of our testing accounts were blocked throughout the study, hinting that we did not cause any significant harm and most likely were not even noticed by the platform operator.
Lastly, to accurately assess the feasibility of the proposed attack, it was essential to conduct experiments against the actual WhatsApp infrastructure.
Given the minimal risk of adverse effects on other users or the service itself, as argued above, we considered this a reasonable approach.
Finally, all our findings have been responsibly disclosed to Meta.

\subsection{Limitations}
Although WhatsApp is the most popular instant messenger using the Signal protocol, many other messaging applications rely on the same protocol suite.
While our analysis specifically focuses on WhatsApp, some of our findings may generalize to other Signal-based messengers.
Due to WhatsApp’s closed-source nature, we were unable to directly inspect the source code of the official clients or the server backend.
However, given its immense popularity, it is crucial to scrutinize its overall security.
We hope this work serves as a first step toward shedding light on WhatsApp’s real-world implementation and the design decisions underlying its deployment of the Signal protocol.

\subsection{Countermeasures}
Many of the exploits and side channels presented in this work are inherent to Signal’s session handshake protocol, which relies on the availability of fresh one-time prekeys.
As a result, completely eliminating these issues is challenging; for example, the device's online state will inevitably be exposed when new prekeys are uploaded.
Nevertheless, we propose several mitigations that 
would substantially reduce the practical exploitability of the identified vulnerabilities.

\parvspace
\noindent
\textbf{Rate Limiting.} A single account should not be able to constantly query prekey bundles for the same device in rapid succession. Given the fast refill rate of most Android devices, even a modest artificial slow down (i.e., rate limiting), would reduce the likelihood of a successful one-time prekey depletion attack against these devices significantly. 

\parvspace
\noindent
\textbf{Reduce Signed Prekey Renewal Interval.} The lifetime of a signed prekey in WhatsApp is higher ($\approx$ month), than the lifetime of a signed prekey in Signal (two days).
Reducing the lifetime of signed prekeys would also reduce the impact regarding PFS through missing one-time prekeys.

\parvspace
\noindent
\textbf{Visual Indication of Missing PFS in the UI.}
Currently neither sender, nor receiver are notified if a new session is established without a one-time prekey, therefore there is no obvious way to detect such an attack as a user.
To not flood all users with complicated warnings and to prevent misunderstandings, the settings could offer a verbose option for security-cautious or high-profile users, which would show such security related UI notifications when messages lack PFS. 

\parvspace
\noindent
\textbf{Signed Prekey Update on Demand.}
If a prekey bundle without a one-time prekey is used to initiate a new session, the responder device could update his signed prekey together with the next batch of one-time prekeys it pushes to the server. 
This would minimize the damage a lack of PFS could cause, in case the responder device is online.
Due to asynchronous communication, there may still exist prekey bundles in circulation that contain outdated signed prekeys, but this should not be a large problem since there is no valid use case for keeping them around and not immediately initiate a new session.
To prevent an attacker from turning this countermeasure into a DoS attack, there should be a minimum validity period (in the order of minutes) for signed prekeys.
Otherwise an attacker could trigger signed prekey updates all the time by initiating new sessions without one-time prekeys, which would prevent everybody else from establishing a new session as signed prekeys are constantly outdated. 

\parvspace
\noindent
\textbf{Redesign Key IDs.} Due to their initial values and the fact that they are incremented by one, key IDs leak information and make device fingerprinting possible. The question is, if key IDs could not be enlarged and entirely be replaced with hashes of the respective public keys they refer to, which would completely mitigate this information disclosure vulnerability. 

\section{Conclusion}

In this work we have demonstrated that WhatsApp does not enforce any rate limiting regarding the querying of prekey bundles, thereby violating the Signal X3DH specification.
This enables an attacker to deplete all one-time prekeys of a targeted device, subsequently degrading the perfect forward secrecy (PFS) of new sessions initiated with the victim. 
Although, PFS is undoubtedly effected by such an attack, the successful exploitation of this degraded forward secrecy would still require a compromise of the involved long- and medium-term secret keys, as well as passive eavesdropping capabilities to record the respective encrypted messages. 

In contrast to this rather strong attacker model, we also describe attacks on privacy and availability, with the sole requirement of having a WhatsApp account. 
Hereby, we were able to show that the refilling of one-time prekeys necessarily leaks the current online status of the respective device, as well as in certain cases: the device age, operating system and the approximate total number of new sessions initiated with the targeted device.
Moreover, we where able to highlight a DoS issue by rapidly querying prekey bundles of a device such that the retrieval of any prekey bundle (even without one-time prekeys) was no longer possible. 
As a consequence, for the duration of the attack no new session can be established with the victim. 
All attacks described in this paper can be executed covertly, and targeted at any of WhatsApp’s more than 3 billion users.

To mitigate the discovered issues, we suggest a range of countermeasures. Most notably the notification of users regarding the degraded PFS in the UI, as well as the introduction of rate limits regarding the repeated fetching of prekey bundles for the same device from a single account.

\section*{Artifact Evaluation}
The artifact accompanying this paper is publicly available at \url{https://github.com/sbaresearch/prekey-pogo/tree/woot25ae}.
It includes a modified WhatsApp client based on \textit{whatsmeow}, which enables the retrieval of a user's WhatsApp devices and their associated prekey material.
As part of the WOOT 2025 artifact evaluation process, we applied for and were awarded both the \textit{Artifacts Available} and \textit{Artifacts Functional} badges.

\section*{Acknowledgments}
This material is based upon work partially supported by
(1) the Christian-Doppler-Laboratory for Security and Quality Improvement in the Production System Lifecycle;
The financial support by the
Austrian Federal Ministry for Digital and Economic Affairs,
the Nation Foundation for Research, Technology and Development
and University of Vienna, Faculty of Computer Science, Security \& Privacy Group
is gratefully acknowledged;
(2) the FFG Bridge project 46322124 SecKey;
(3) SBA Research (SBA-K1 NGC), a COMET Center within the COMET – Competence Centers for Excellent Technologies Programme, funded by BMIMI, BMWET, and the federal state of Vienna. The COMET Programme is managed by FFG.

We further would like to thank Markus Maier for his outstanding support in providing and maintaining the IT infrastructure, which was essential for the success of this work.
Moreover, we would also like to thank our anonymous reviewers for their valuable feedback and suggestions.

\pagebreak

\bibliography{bibliography,background}

\begin{thebibliography}{10}

\bibitem{noauthor_crystals_nodate}
{CRYSTALS} cryptographic suite for algebraic lattices.
\newblock Retrieved Aug 26th, 2024 from
  \url{https://pq-crystals.org/index.shtml}.

\bibitem{noauthor_whatsapp_2023}
{WhatsApp} encryption overview: Technical white paper.
\newblock Retrieved Aug 26th, 2024 from
  \url{https://faq.whatsapp.com/82012443585354}.

\bibitem{alwen_double_2019}
Joël Alwen, Sandro Coretti, and Yevgeniy Dodis.
\newblock The double ratchet: Security notions, proofs, and modularization for
  the signal protocol.
\newblock In Yuval Ishai and Vincent Rijmen, editors, {\em Advances in
  Cryptology - {EUROCRYPT} 2019 - 38th Annual International Conference on the
  Theory and Applications of Cryptographic Techniques, Darmstadt, Germany, May
  19-23, 2019, Proceedings, Part I}, volume 11476 of {\em Lecture Notes in
  Computer Science}, pages 129--158. Springer.

\bibitem{beery_whatsapp_2024}
Tal~A. Be{\textquoteright}ery.
\newblock {WhatsApp} with privacy? privacy issues with {IM} e2ee in the
  multi-device setting.
\newblock In {\em 18th {USENIX} {WOOT} Conference on Offensive Technologies
  ({WOOT} 24)}, pages 11--16. {USENIX} Association.

\bibitem{Curve25519}
Daniel~J. Bernstein.
\newblock Curve25519: New diffie-hellman speed records.
\newblock In Moti Yung, Yevgeniy Dodis, Aggelos Kiayias, and Tal Malkin,
  editors, {\em Public Key Cryptography - PKC 2006}, pages 207--228, Berlin,
  Heidelberg, 2006. Springer Berlin Heidelberg.

\bibitem{brendel_post-quantum_2022}
Jacqueline Brendel, Rune Fiedler, Felix Günther, Christian Janson, and Douglas
  Stebila.
\newblock Post-quantum asynchronous deniable key exchange and the signal
  handshake.
\newblock In Goichiro Hanaoka, Junji Shikata, and Yohei Watanabe, editors, {\em
  Public-Key Cryptography - {PKC} 2022 - 25th {IACR} International Conference
  on Practice and Theory of Public-Key Cryptography, Virtual Event, March 8-11,
  2022, Proceedings, Part {II}}, volume 13178 of {\em Lecture Notes in Computer
  Science}, pages 3--34. Springer.

\bibitem{chase_signal_2020}
Melissa Chase, Trevor Perrin, and Greg Zaverucha.
\newblock The signal private group system and anonymous credentials supporting
  efficient verifiable encryption.
\newblock In Jay Ligatti, Xinming Ou, Jonathan Katz, and Giovanni Vigna,
  editors, {\em {CCS} '20: 2020 {ACM} {SIGSAC} Conference on Computer and
  Communications Security, Virtual Event, {USA}, November 9-13, 2020}, pages
  1445--1459. {ACM}.

\bibitem{private_groups}
Melissa Chase, Trevor Perrin, and Greg Zaverucha.
\newblock {The Signal Private Group System and Anonymous Credentials Supporting
  Efficient Verifiable Encryption}.
\newblock In Jay Ligatti, Xinming Ou, Jonathan Katz, and Giovanni Vigna,
  editors, {\em {CCS} '20: 2020 {ACM} {SIGSAC} Conference on Computer and
  Communications Security, Virtual Event, USA, November 9-13, 2020}, pages
  1445--1459. {ACM}, 2020.

\bibitem{cohn-gordon_formal_2020}
Katriel Cohn-Gordon, Cas Cremers, Benjamin Dowling, Luke Garratt, and Douglas
  Stebila.
\newblock A formal security analysis of the signal messaging protocol.
\newblock 33(4):1914--1983.

\bibitem{cohn-gordon_post-compromise_2016}
Katriel Cohn-Gordon, Cas Cremers, and Luke Garratt.
\newblock On post-compromise security.
\newblock In {\em {IEEE} 29th Computer Security Foundations Symposium, {CSF}
  2016, Lisbon, Portugal, June 27 - July 1, 2016}, pages 164--178. {IEEE}
  Computer Society.

\bibitem{clone2020}
Cas Cremers, Jaiden Fairoze, Benjamin Kiesl, and Aurora Naska.
\newblock {Clone Detection in Secure Messaging: Improving Post-Compromise
  Security in Practice}.
\newblock In Jay Ligatti, Xinming Ou, Jonathan Katz, and Giovanni Vigna,
  editors, {\em {CCS} '20: 2020 {ACM} {SIGSAC} Conference on Computer and
  Communications Security, Virtual Event, USA, November 9-13, 2020}, pages
  1481--1495. {ACM}, 2020.

\bibitem{cremers_formal_2023}
Cas Cremers, Charlie Jacomme, and Aurora Naska.
\newblock {Formal Analysis of Session-Handling in Secure Messaging: Lifting
  Security from Sessions to Conversations}.
\newblock In Joseph~A. Calandrino and Carmela Troncoso, editors, {\em 32nd
  {USENIX} Security Symposium, {USENIX} Security 2023, Anaheim, {CA}, {USA},
  August 9-11, 2023}, pages 1235--1252. {USENIX} Association.

\bibitem{freed_stalker_2018}
Diana Freed, Jackeline Palmer, Diana Minchala, Karen Levy, Thomas Ristenpart,
  and Nicola Dell.
\newblock {“A Stalker's Paradise” How Intimate Partner Abusers Exploit
  Technology}.
\newblock In {\em Proceedings of the 2018 CHI conference on human factors in
  computing systems}, pages 1--13, 2018.

\bibitem{gegenhuber_2025_heythere}
Gabriel~K. Gegenhuber, Philipp~É. Frenzel, Maximilian Günther, Johanna
  Ullrich, and Aljosha Judmayer.
\newblock {Hey there! You are using WhatsApp: Enumerating Three Billion
  Accounts for Security and Privacy}, 2025.

\bibitem{gegenhuber_2024_carelesswhisper}
Gabriel~K. Gegenhuber, Maximilian Günther, Markus Maier, Aljosha Judmayer,
  Florian Holzbauer, Philipp~É. Frenzel, and Johanna Ullrich.
\newblock {Careless Whisper: Exploiting Silent Delivery Receipts to Monitor
  Users on Mobile Instant Messengers}, 2024.

\bibitem{gegenhuber_2024_diffie}
Gabriel~K. Gegenhuber, Florian Holzbauer, Philipp~{\'E}. Frenzel, Edgar Weippl,
  and Adrian Dabrowski.
\newblock {{Diffie-Hellman} Picture Show: Key Exchange Stories from Commercial
  {VoWiFi} Deployments}.
\newblock In {\em 33rd USENIX Security Symposium (USENIX Security 24)}, pages
  451--468, Philadelphia, PA, August 2024. USENIX Association.

\bibitem{gegenhuber_2023_mobileatlas}
Gabriel~K. Gegenhuber, Wilfried Mayer, Edgar Weippl, and Adrian Dabrowski.
\newblock {MobileAtlas: Geographically Decoupled Measurements in Cellular
  Networks for Security and Privacy Research}.
\newblock In {\em Usenix Security Symposium 2023}, 2023.

\bibitem{hagen_2021_all}
Christoph Hagen, Christian Weinert, Christoph Sendner, Alexandra Dmitrienko,
  and Thomas Schneider.
\newblock {All the numbers are US: Large-scale abuse of contact discovery in
  mobile messenger}.
\newblock In {\em 28th Annual Network and Distributed System Security
  Symposium, {NDSS} 2012, San Diego, California, {USA}, February 21 - February
  25, 2021}. The Internet Society.

\bibitem{kret_pqxdh_2023}
Ehren Kret and Rolfe Schmidt.
\newblock The {PQXDH} key agreement protocol.
\newblock Retrieved Aug 26th, 2024 from
  \url{https://signal.org/docs/specifications/pqxdh/pqxdh.pdf}.

\bibitem{marlinspike_private_2014}
Moxie Marlinspike.
\newblock Private group messaging.
\newblock Retrieved Aug 26th, 2024 from
  \url{https://signal.org/blog/private-groups/}.

\bibitem{marlinspike_sesame_2017}
Moxie Marlinspike and Trevor Perrin.
\newblock The sesame algorithm: Session management for asynchronous message
  encryption.
\newblock Retrieved Aug 26th, 2024 from
  \url{https://signal.org/docs/specifications/sesame/sesame.pdf}.

\bibitem{marlinspike_x3dh_2016}
Moxie Marlinspike and Trevor Perrin.
\newblock The x3dh key agreement protocol.
\newblock Retrieved Aug 26th, 2024 from
  \url{https://signal.org/docs/specifications/x3dh/x3dh.pdf}.

\bibitem{meta2023messenger}
Meta.
\newblock Messenger end-to-end encryption overview, December 2023.
\newblock Accessed: 2025-03-11.

\bibitem{mueller_2014_what}
Robin Mueller, Sebastian Schrittwieser, Peter Fruehwirt, Peter Kieseberg, and
  Edgar Weippl.
\newblock {What's new with WhatsApp \& Co.? Revisiting the Security of
  Smartphone Messaging Applications}.
\newblock iiWAS '14, page 142–151, New York, NY, USA, 2014. Association for
  Computing Machinery.

\bibitem{NIST800-56A-R3}
{National Institute of Standards and Technology}.
\newblock {Recommendation for Pair-Wise Key-Establishment Schemes Using
  Discrete Logarithm Cryptography (Revision 3)}.
\newblock Technical Report NIST SP 800-56A Rev. 3, National Institute of
  Standards and Technology, April 2018.

\bibitem{perrin2018noise}
Trevor Perrin.
\newblock {The Noise Protocol Framework}.
\newblock Online, 2018.
\newblock Version 34, Retrieved March 5, 2025.

\bibitem{perrin_double_2016}
Trevor Perrin and Moxie Marlinspike.
\newblock The double ratchet algorithm.
\newblock Retrieved Aug 26th, 2024 from
  \url{https://signal.org/docs/specifications/doubleratchet/doubleratchet.pdf}.

\bibitem{schnitzler_hope_2023}
Theodor Schnitzler, Katharina Kohls, Evangelos Bitsikas, and Christina Pöpper.
\newblock Hope of delivery: Extracting user locations from mobile instant
  messengers.
\newblock In {\em 30th Annual Network and Distributed System Security
  Symposium, {NDSS} 2023, San Diego, California, {USA}, February 27 - March 3,
  2023}. The Internet Society.

\bibitem{schrittwieser_2012_guess}
Sebastian Schrittwieser, Peter Frühwirt, Peter Kieseberg, Manuel Leithner,
  Martin Mulazzani, Markus Huber, and Edgar Weipp.
\newblock {Guess who is texting you? Evaluating the security of smartphone
  messaging applications}.
\newblock In {\em 19th Annual Network and Distributed System Security
  Symposium, {NDSS} 2012, San Diego, California, {USA}, February 5 - February
  8, 2012}. The Internet Society.

\bibitem{tu_2016_new}
Guan-Hua Tu, Chi-Yu Li, Chunyi Peng, Yuanjie Li, and Songwu Lu.
\newblock {New security threats caused by IMS-based SMS service in 4G LTE
  networks}.
\newblock In {\em Proceedings of the 2016 ACM SIGSAC Conference on Computer and
  Communications Security}, 2016.

\bibitem{noauthor_whatsapp_about}
WhatsApp.
\newblock {About WhatsApp}, 2024.
\newblock Retrieved Aug 26th, 2024 from \url{https://www.whatsapp.com/about/}.

\bibitem{wichelmann_help_2021}
Jan Wichelmann, Sebastian Berndt, Claudius Pott, and Thomas Eisenbarth.
\newblock {Help, My Signal has Bad Device! - Breaking the Signal Messenger's
  Post-Compromise Security Through a Malicious Device}.
\newblock In Leyla Bilge, Lorenzo Cavallaro, Giancarlo Pellegrino, and Nuno
  Neves, editors, {\em Detection of Intrusions and Malware, and Vulnerability
  Assessment - 18th International Conference, {DIMVA} 2021, Virtual Event, July
  14-16, 2021, Proceedings}, volume 12756 of {\em Lecture Notes in Computer
  Science}, pages 88--105. Springer.

\bibitem{yang_2024_uncovering}
Yaru Yang, Yiming Zhang, Tao Wan, Chuhan Wang, Haixin Duan, Jianjun Chen, and
  Yishen Li.
\newblock {Uncovering Security Vulnerabilities in Real-world Implementation and
  Deployment of 5G Messaging Services}.
\newblock In {\em Proceedings of the 17th ACM Conference on Security and
  Privacy in Wireless and Mobile Networks}, 2024.

\end{thebibliography}
\bibliographystyle{plain}

\appendix

\pagebreak
\renewcommand{\thesubsection}{\Alph{subsection}}
\section*{Appendix}
\label{sec:appendix}

\subsection{Tested Phone Models and OS Versions}

\begin{table*}[b]
    \centering
    \begin{tabular}{@{}lllll@{}}
    \toprule
    Device                & Modem Chipset & OS                       & WhatsApp   \\ \midrule
    iPhone SE 2020        & Intel         & iOS 18.3.1               & 2.25.4.77 \\
    iPhone 8              & Intel         & iOS 16.7.10              & 2.25.5.74 \\
    iPhone 11             & Qualcomm      & iOS 18.3.1               & 2.25.5.74 \\
    Xiaomi POCO X3 NFC    & Qualcomm      & Android 12 (MIUI 14.0.5) & 2.25.2.78 \\
    Samsung Galaxy A54 5G & Exynos        & Android 14               & 2.25.2.78 \\
    Xiaomi Redmi 10 5G    & MediaTek      & Android 14               & 2.25.2.78 \\
    \bottomrule
    \end{tabular}
    \caption{Overview of the devices including software versions that were used throughout our tests.
    For our WhatsApp Web tests (Chrome, Firefox, Safari) we've used the most recent browser and WhatsApp web versions available (testing date 2025-02-21).
    }
    \label{tab:testing-devices}
\end{table*}

\begin{figure*}[h]
    \centering    
        \includegraphics[width=0.95\textwidth]{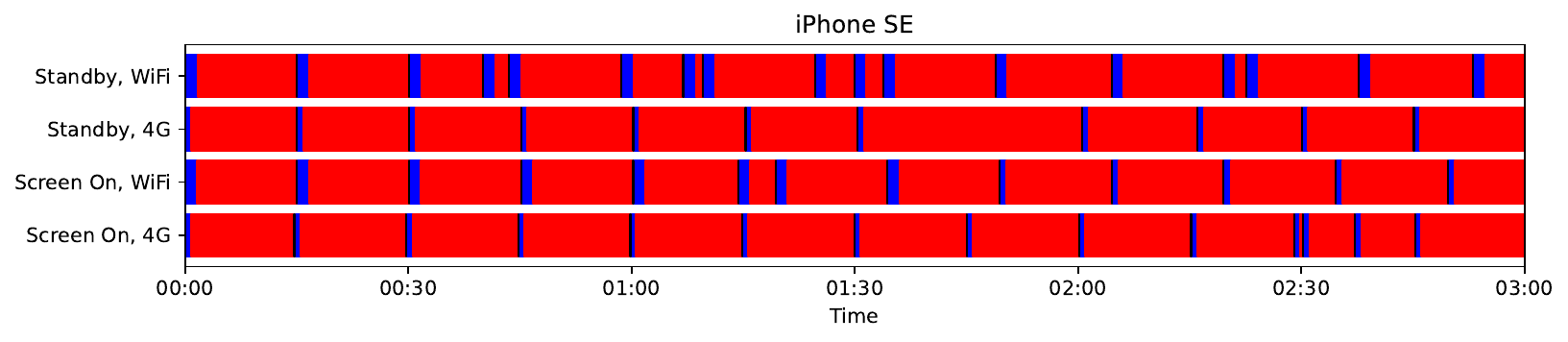}
        \includegraphics[width=0.95\textwidth]{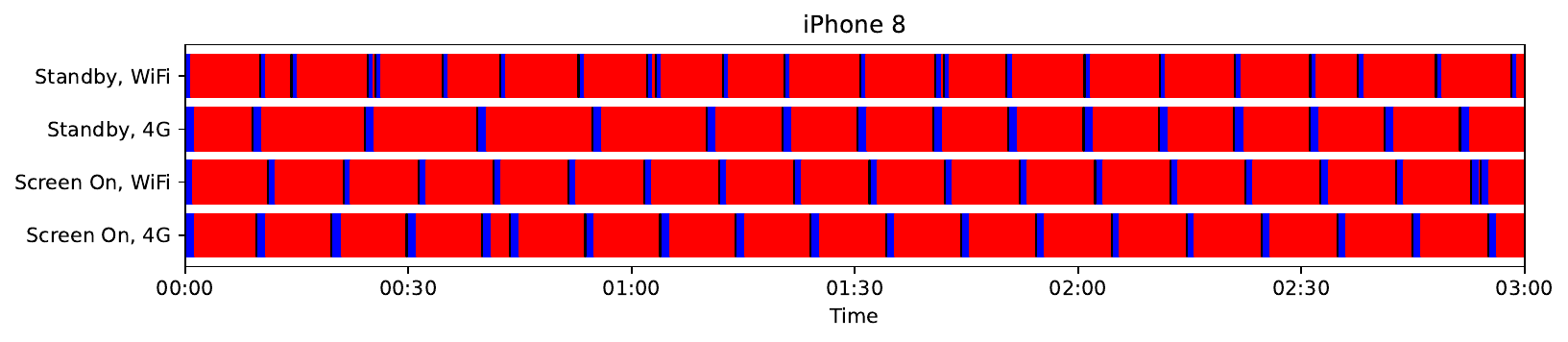}
        \includegraphics[width=0.95\textwidth]{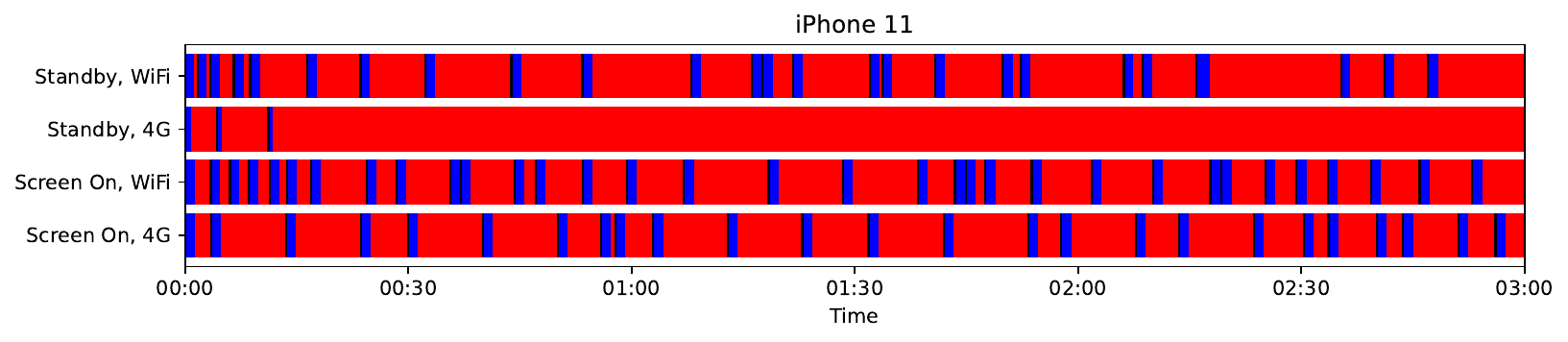}
        \includegraphics[width=0.95\textwidth]{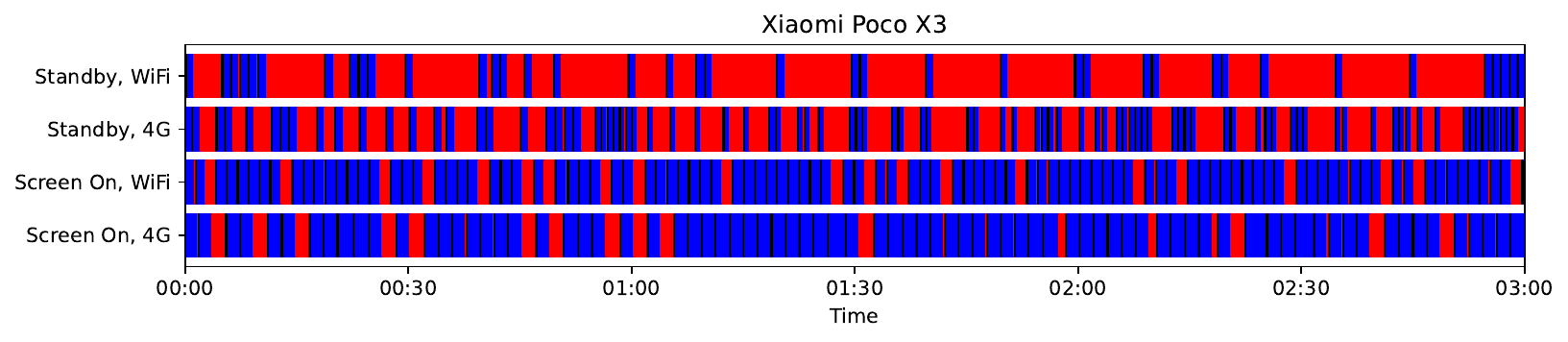}
        \includegraphics[width=0.95\textwidth]{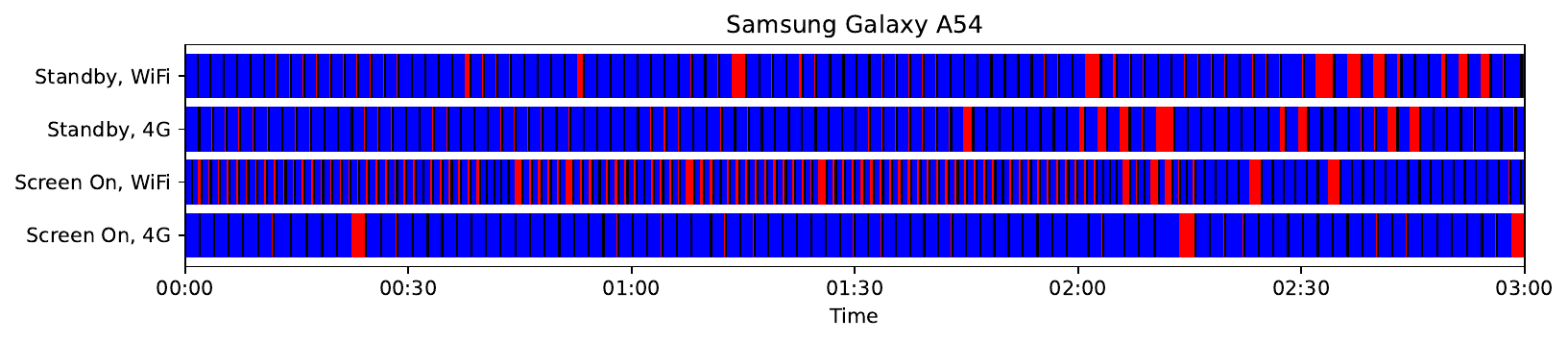}
        \includegraphics[width=0.95\textwidth]{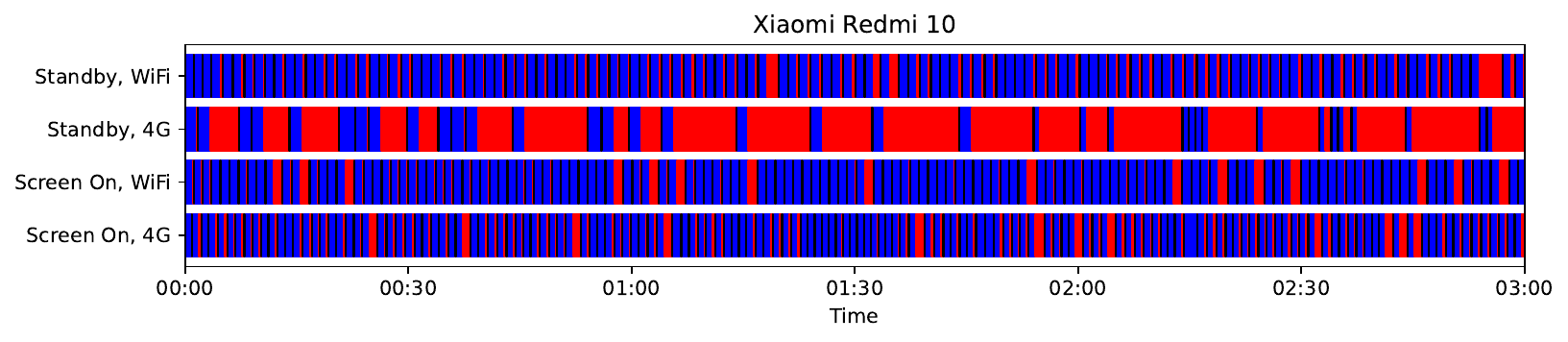}
        \caption{Characteristic refill behavior of different smartphone models in various device and connection states. Time where no one-time prekeys are available is shown in {\color{red}red}. }
        \label{fig:characteristic-device-behaviour}
\end{figure*}

\subsection{Signal Protocol in More Detail}
\label{sec:signal}

The Signal protocol actually consists of an entire family of protocols~\cite{marlinspike_x3dh_2016, perrin_double_2016, marlinspike_private_2014, marlinspike_sesame_2017, kret_pqxdh_2023} which been studied in a variety of works~\cite{cohn-gordon_post-compromise_2016, alwen_double_2019, cohn-gordon_formal_2020, brendel_post-quantum_2022,wichelmann_help_2021, cremers_formal_2023, clone2020}. 

Internally, the signal protocol uses two key derivation functions (KDF), one to derive ratchet keys, which we denote by $KDF_r(\cdot)$ and one to derive the message keys, denoted by $KDF_m(\cdot)$. The KDFs are implemented with HKDF and HMAC-SHA256. The thereby generated derived keys are used for authenticated encryption with associated data (AEAD), using AES256 in CBC mode for encryption and HMAC-SHA256 for authentication. We denote the symmetric encryption function by $E(\text{key},\text{plaintext},\text{associated data})$ and the decryption function by $D(\text{key},\text{ciphertext},\text{associated data})$. Moreover, the signal protocol relies on Diffie-Hellman key exchange to compute shared keys and achieve its design goals. We denote the key exchange algorithm by $DH(\cdot)$, which is implemented over Curve25519~\cite{Curve25519} in practice

The Signal protocol uses three different types of Diffi-Hellman public keys to ensure forward secrecy right from the start: 
Long-term identity keys, medium-term (signed) prekeys and short-term (one-time) ephemeral prekeys (see Table~\ref{tab:keys} for an overview). 
In our scenario Alice is the \emph{initiator} and wants to establish a secure connection with Bob (the \emph{responder}). 
The Signal protocol, as also implemented by WhatsApp, uses prekey bundles deposited by every user at a central server to allow any \emph{initiator} to negotiate a shared secret (via Diffi-Hellman Key exchange) even if the \emph{responder} is currently not online/available using the X3DH protocol~\cite{marlinspike_x3dh_2016}~\footnote{Since late 2023 Signal replaced X3DH by PQXDH~\cite{kret_pqxdh_2023} and started a process to use PQXDH for new sessions if supported by both peers. On a high level, PQ3DH is comparable to X3DH with the difference that an additional key from a CRYSTALS-Kyber~\cite{noauthor_crystals_nodate} key encapsulation mechanism (KEM) is used in the KDF.}. 

\begin{table*}[]
\centering
\begin{tabular}{@{}lll@{}}
\toprule
Keys    &                 & Description                          \\ \midrule
$ipk^A$ & $ik^A $         & Long-term identity key pair of Alice \\
$ipk^B $ & \color{orange} $ik^B $      & Long-term identity key pair of Bob   \\
$prepk^A$ & $ prek^A  $  & Medium-term prekey pair of Alice, aka. \emph{signed prekey} \\ 
$prepk^B$ & \color{orange} $prek^B  $  & Medium-term prekey pair of Bob, aka. \emph{signed prekey} \\ 
$eprepk^A_n$ & $eprek^A_n $ & Short-term prekey pair number $n$ of Alice, aka. \emph{ephemeral prekey} or \emph{one-time prekey} \\
$eprepk^B_n$ & $eprek^B_n $ & Short-term prekey pair number $n$ of Bob, aka. \emph{ephemeral prekey} or \emph{one-time prekey} \\  \midrule
\multicolumn{2}{l}{$\langle ipk^A, prepk^A, Sig(ik^A, prepk^A), {\color{red}[ eprek^A_n ]}\rangle$} & A prekey bundle deposited by Alice on the server \\
\multicolumn{2}{l}{$\langle ipk^B, prepk^B, Sig(ik^B, prepk^B), {\color{red}[ eprek^B_n ]}\rangle$} & A prekey bundle deposited by Bob on the server \\ \midrule
$ epk^A$  & $ ek^A $     & Ephemeral handshake key pair of Alice \\ 
$ rchpk_0^A$ & $rchk_0^A $  & Ephemeral ratchet key pair of Alice \\ \midrule
\multicolumn{2}{l}{$ rk_x $ }                 & Symmetric (shared) root key of ratchet number $x$ \\
\multicolumn{2}{l}{$ ck^{i \to r}_{x,y} $ }   & Symmetric (shared) chaining key number $y$, in the $x^{th}$ initiator to responder ratchet \\
\multicolumn{2}{l}{$ mk^{i \to r}_{x,y} $ }   & Symmetric (shared) message key number $y$, in the $x^{th}$ initiator to responder ratchet \\ 
\bottomrule
\end{tabular}
\caption{Main cryptographic keys of the signal protocol relevant for our attacks. Public keys in asymmetric schemes always end in $pk$. The naming convention of the keying material is according to Cohn-Gordon et al.~\cite{cohn-gordon_formal_2020}. The ephemeral prekeys, which are considered optional in the prekey bundle, are depicted in {\color{red}red}. The secret keys of Bob which an attacker has to compromise to benefit from the violation of forward secrecy, i.e., if no ephemeral prekeys can be used, are depicted in {\color{orange}orange}.}
\label{tab:keys}
\end{table*}

The initial handshake works as depicted in figure~\ref{fig:no-prekey} in the appendix and here starting with formula~\ref{for:handshake}. First the prekey bundle of the responder (in our case Bob) is fetched from the server by the initiator (in our case Alice). 
The information from the prekey bundle is verified by Alice through checking the signature on the \emph{signed prekey} using the (long-term) identity pubic key of Bob. As within other works~\cite{cohn-gordon_formal_2020}, it is assumed that Alice has already verified out-of-band that the long-term identity public key indeed belongs to Bob. 

Then the public keys from Bob's prekey bundle are used to compute shared keys for the ratcheting and message encryption and authentication. Here now we have to distinguish between the case where an ephemeral (one-time) prekey $eprepk^B$ of Bob is available or not. If no ephemeral prekey is available, the DH invocation $dh_4$ in formula~\ref{for:dh4} is omitted. 

In any case, before initiating the session and sending the first message to Bob, Alice generates two ephemeral key pairs: The \emph{ephemeral handshake key pair} denoted $(epk^A, ek^A)$ and the \emph{ephemeral ratchet key pair} denoted $(rchpk^A, rchk^A)$. Those are used for the initial handshake and to initialize the DH ratchet construction, also referred to as the \emph{asymmetric ratchet}. 

\begin{align}
\label{for:handshake}
dh_1  &\gets DH(ik^A,prepk^B) \\
dh_2  &\gets DH(ek^A,ipk^B)\\
dh_3  &\gets DH(ek^A,prepk^B)\\
\color{red} [dh_4 &\color{red} \gets  DH(ek^A,eprepk^B) ] \\
\label{for:dh4}
rk_0 &\gets KDF_{r}(dh_1 \mid\mid dh_2 \mid\mid dh_3 {\color{red}[ \mid\mid DH4 ]})\\
DH_{ratchet} &\gets DH(rchk_0^A, prepk^B) \\
rk_1,ck^{i \to r}_{0,0} &\gets KDF_{r}( rk_0,  DH_{ratchet} )\\
{\color{Purple}ck_{0,1}^{i \to r}}, mk_{0,0}^{i \to r} &\gets KDF_{m}(ck^{i \to r}_{0,0})
\end{align}

The derived message key $mk_{0,0}^{i \to r}$ is then used to encrypt and authenticate ($AE$) the first chat \emph{message} from Alice to Bob, as well as to authenticate some associated data $AD$ consisting of the ephemeral public keys ($ epk^A $ and $ rchpk_0^A$ ) generated previously by Alice. 

\begin{align}
\label{for:init}
AD &\gets \langle rchpk_0^A, epk^A, \text{ id of } prepk^B, {\color{red} [ eprepk^B ] } \rangle \\
AE_{mk_{0,0}^{i \to r}}, AD &\gets E(mk_{0,0}^{i \to r},message,AD)
\end{align}

Once Bob receives this initial message, he can compute the same shared keys using his identity key ${\color{orange}ik^B}$ and his prekey ${\color{orange}prek^B}$, as well as the public keys of Alice consisting of her identity key $ipk^A$, her ephemeral handshake key $epk^A$ and her ephemeral ratchet key $rchpk^A$. 
Since the later two are transmitted in the associated data $AD$, they are authenticated, but not encrypted. 

\begin{align}
dh_1 &= DH(ipk^A,{\color{orange}prek^B}) \\
dh_2 &= DH(epk^A,{\color{orange}ik^B})\\
dh_3 &= DH(epk^A,{\color{orange}prek^B})\\
rk_0 &\gets KDF_{r}(dh_1 \mid\mid dh_2 \mid\mid dh_3)\\
DH_{ratchet} &\gets DH(rchpk_0^A, {\color{orange}prek^B}) \\
rk_1,ck^{i \to r}_{0,0} &\gets KDF_{r}(rk_0, DH_{ratchet} )\\
{\color{Purple}ck_{0,1}^{i \to r}}, mk_{0,0}^{i \to r} &\gets KDF_{m}(ck^{i \to r}_{0,0})
\end{align}

The received message is then decrypted using the previously computed shared message key $mk_{0,0}^{i \to r}$:
\begin{align}
    message,AD &\gets D(mk_{0,0}^{i \to r},AE_{mk_{0,0}^{i \to r}}, AD)
\end{align}

If no ephemeral prekeys have been fetched by Alice, this initial message sent by Alice has no forward secrecy if observed by an attacker. 
Therefore, an attacker who is able to compromise Bobs medium-term and long-term secret keys $prek^B$ and $ik^B$ later on, can recompute the same message key $mk_{0,0}^{i \to r}$, which highlights that there is no forward secrecy for this message. 
If Alice sends multiple messages before receiving any response from Bob, all these messages are affected as well, as the keys for these messages come from the symmetric ratchet. This is illustrated by the following example starting with formula~\ref{for:multi}, which depicts encrypting a second message from Alice to Bob. Here, $y$ is $0$ at the beginning and later set to $y=1$ for the second message and so forth:

\begin{align}
\label{for:multi}
ck_{0,2}^{i \to r}, mk_{0,1}^{i \to r} &\gets KDF_{m}({\color{Purple}ck_{0,1}^{i \to r}})\\
y &\gets y+1 \\
AD &\gets (rchpk_0^A,ipk^A,ipk^B,y)\\
AE_{mk_{0,1}^{i \to r}},AD &\gets AE(mk_{0,1}^{i \to r},message,AD)
\end{align}

Forward secrecy is restored through the asymmetric ratchet, when Bob responds to a message. 
If Bob responds to one of Alice messages, he also computes a new ephemeral ratchet key pair ($rchpk^B_1, rchk^B_1$), s.t. $x=1$, and thereby advances the asymmetric ratchet as follows:

\begin{align}
    DH_{ratchet} &\gets DH(rchpk_{x-1}^A, rchk^B_x) \\
    tmp,ck^{r \to i}_{x,0} &\gets KDF_{r}(rk_x, DH_{ratchet} )\\
    ck_{x,1}^{r \to i}, mk_{x,0}^{r \to i} &\gets KDF_{m}(ck^{r \to i}_{x,0})\\
    AD &\gets (rchpk_x^B)\\
    AE_{mk_{x,y}^{r \to i}} &\gets E(mk_{x,y}^{r \to i},message,AD)
\end{align}

As soon as these ephemeral ratchet keys are deleted, forward secrecy is regained for this as well as subsequent messages.
Note, that even if forward secrecy is regained in a chat session, the initial messages sent from Alice to Bob have been encrypted using the symmetric ratchet only.
Therefore, they remain vulnerable for the entire lifetime of the signed prekey $prek^B$ of Bob.
According to the specifications, the signed prekey should be periodically rotated~\cite{noauthor_whatsapp_2023,marlinspike_x3dh_2016,meta2023messenger}, where suggested intervals reach from once a week to once a month~\footnote{In practice Signal rotates the signed prekey every two days~\url{https://github.com/signalapp/Signal-Android/blob/481dc162d80292a046b4229cceba2ac2f2a73f36/app/src/main/java/org/thoughtcrime/securesms/jobs/PreKeysSyncJob.kt\#L57-L66}}.

\begin{figure*}
    \centering
    \includegraphics[width=0.90\linewidth]{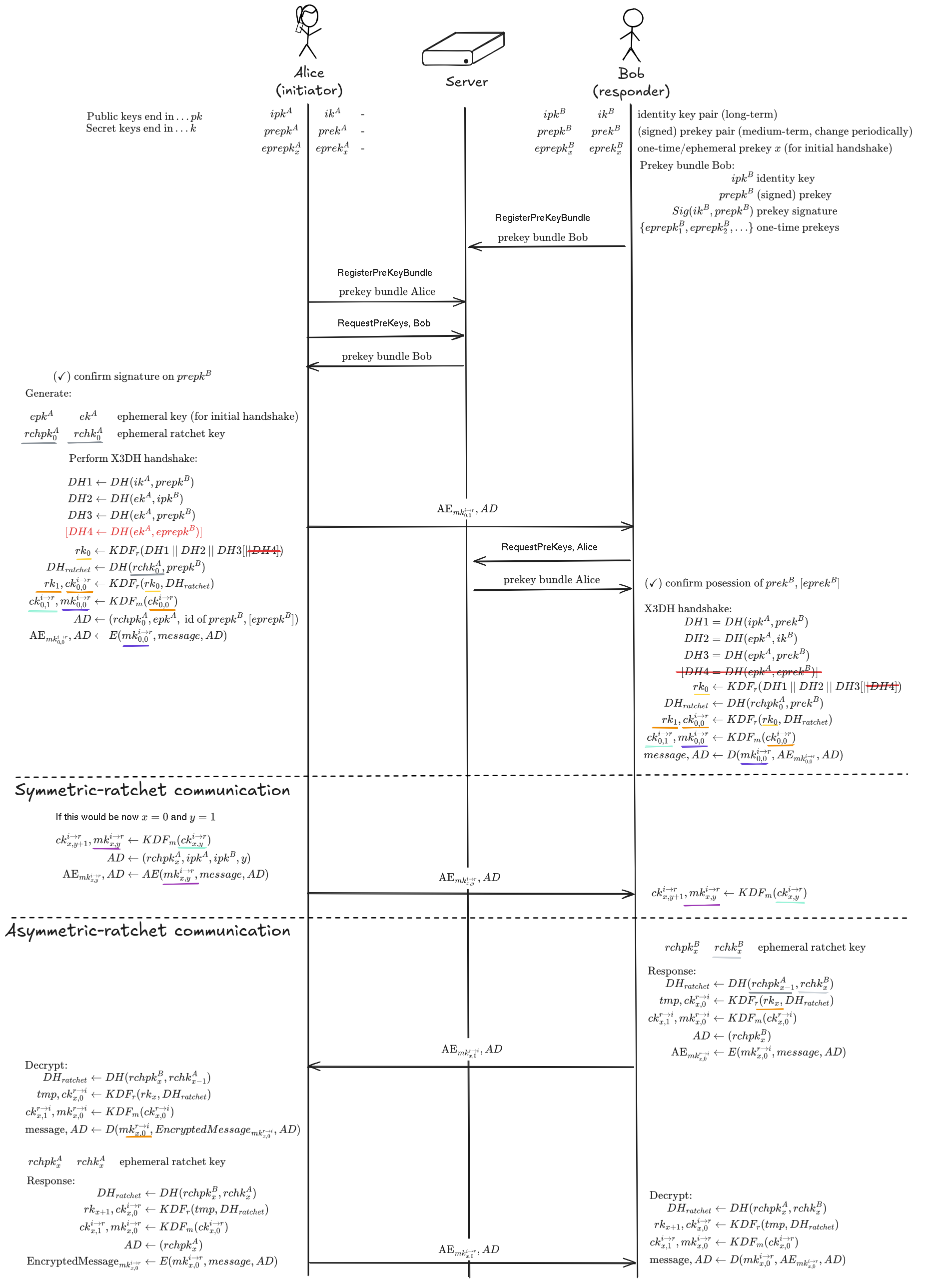}
    \caption{Signal protocol layout, when no \emph{ephemeral (one-time) prekeys} are available on the server. Identical keys a highlighted with the same color. }
    \label{fig:no-prekey}
\end{figure*}

\end{document}